\documentclass[12pt]{article}

\usepackage{epsf}
\usepackage[dcu]{harvard}

\newcommand{\eq}{\begin{equation}}
\newcommand{\en}{\end{equation}}
\newcommand{\bd}{\begin{displaymath}}
\newcommand{\ed}{\end{displaymath}}
\newcommand{\minimo}[2]{\mbox{\emph{min}$_{_{_{\hspace*{-11pt}#1\hspace*{6pt}}}}\{#2\}$}}
\newcommand{\ignorar}[1]{}

\title{ {\bf 
The multi-fractal structure of contrast changes in natural images:
from sharp edges to textures.
} }

\author{
{\bf\sc Antonio Turiel}\thanks{E-mail: amturiel@delta.ft.uam.es}
 and {\bf\sc N{\'e}stor Parga}\thanks{To whom correspondence should be 
addressed. E-mail: parga@delta.ft.uam.es}
\\
\it Departamento de F\'{\i}sica Te\'orica \\
\it Universidad Aut\'onoma de Madrid \\
\it Canto Blanco, 28049 Madrid, Spain
}

\begin{document}

\maketitle 

\begin{abstract}
We present a formalism that leads very naturally to a hierarchical
description of the different contrast structures in images, providing
precise definitions of sharp edges and other texture components.
Within this formalism, we achieve a decomposition of pixels of the
image in sets, the fractal components of the image, such that each set
only contains points characterized by a fixed stregth of the
singularity of the contrast gradient in its neighborhood.  A crucial
role in this description of images is played by the behavior of
contrast differences under changes in scale.  Contrary to naive
scaling ideas where the image is thought to have uniform
transformation properties \cite{Fie87}, each of these fractal
components has its own transformation law and scaling exponents.  A
conjecture on their biological relevance is also given.

\end{abstract}

{\it Neural Computation} {\bf 12}, 763-793 (2000)

\newpage

\section{Introduction}
\label{section:introduction}


\indent 
It has been frequently remarked that natural images contain redundant
information. Two nearby points in a given image are very likely to
have very similar values of light intensity, except when an edge lies
between them. In this case, the sharp change in luminosity also
represents a loss of predictability, because probably the two points
belong to two different objects, with almost independent luminosities.
Since they cannot be predicted easily, it is frequently said that
edges are the most informative structures in the image and that they
represent their independent features \cite{Bel+97}. Anyone{\'{}}s
subjective experience is consistent with this: most people would start
drawing a scene by tracing the lines of the obvious contours in the
scene. Even more, the picture can be understood by anyone observing
that sketch. Only afterwards, one starts adding texture to the
light-flat areas in the drawing. These textures represent the less
informative data about the position and intensity of the light
sources. It seems that different structures can be distinguished in
every image, following a hierarchy defined according to their
information content, from the sharpest edges to the softest textures.

\indent 
Image processing techniques also emphasize the fact that edges are the
most prominent structures in a scene.  A large variety of methods for
edge detection have been proposed, e.g. high pass filtering,
zero-crossing, wavelet projections, ... \cite{Gon+92}. But once the
edges have been extracted from the image, these methods do not address
the problem of how to describe and obtain the other structures.  In
order to detect these other sets and to extract them in the context of
an organized hierarchy, it is necessary to provide a more general
mathematical framework and, in particular, to define the variables
suitable to perform this classification.

\indent
In this paper we present a more rigorous approach to these issues,
proposing precise definitions of edges and other texture components of
images. As we shall see, our formalism leads very naturally to a
hierarchical description of the different structures in the images,
achieving a decomposition of its pixels in sets, or components,
ranging from the most to the less informative structures.  

\indent 
Another motivation to understand the role played by edges in natural
images is related to the development of the early visual system.  The
brain has to solve the problem of how to represent and transmit
efficiently the information it receives from the outside world.  As it has
been frequently pointed out \cite{Bar61,Lin88,Ati92,vanH92}, cells
located in the first stages of the visual pathway do this by trying to
take advantage of statistical regularities in the image ensemble.  The
idea is that since natural images are highly redundant over space, they
have to be represented in a compact and non-redundant way before being
transmitted to inner areas in the brain.  In order to do this, it is first
necessary to know the regularities of the images, which then could be used
to build more appropriate and efficient internal representations of the
environment. 

\indent
The statistical properties of natural images have been studied for
many years, but the effort was mainly restricted to the second order
statistics. The emphasis was put on the two-point correlation first
because of the simplicity of the analysis, and secondly because its
Fourier transform (i.e. the power spectrum) has a power law behavior
that reflects the existence of scale invariant properties in natural
images.  This implies that the second order statistics does capture
some of the regularities.  However most of the underlying statistical
structure remains unveiled. One can easily convince oneself that most
of the correlations in the image are still present after whitening is
performed: the contours of the objects in the whitened image are still
easily recognizable \cite{Fie87}.  The conclusion is that non-gaussian
statistical properties of edges are important to describe natural
images, a fact that has also been recognized by
\cite{Rud+94,Rud94}. However, no systematic studies of the
statistical properties of edges had been done until very recently
\cite{Tur+98}.

\indent
A complete characterization of the regularities of natural images is
impossible, and probably useless.  Some intuition is necessary to
guide the search for regularities.  The aim of this work is to exploit
the basic idea that changes in contrast are relevant because they
occur at the most informative pixels of the scene. Hence we focus our
attention in the statistics of changes in luminosity.  Since these
changes are graded, the task could seem rather difficult and even
hopeless, but the solution to this problem turns out to be quite
simple and elegant. A single parameter is enough to describe a full
hierarchy of differences in contrast and the hierarchy can be
explicitly constructed. Moreover, it is intimately related to the
scale invariance properties of the image, which are certainly more
elaborated than those that have been used until now but are still
simple enough and have plenty of structure to be detected by a network
adapting to the statistics of the stimuli.


\indent 
To understand these scale invariances we need first the basic concepts
of multiscaling and multifractality \cite{Fal90}, together with a simple
explanation of their meaning for images. While in naive scaling there
is a single transformation law throughout the image (which is then
said to be a fractal), in the more general case there is no invariance
under a global scale transformation but the space can be decomposed in
sets with a different transformation law for each of them (the image
is then said to be a multifractal).

\indent 
To make this concept more precise, let us consider a simple
situation where a set of images present a single type of changes in
contrast: the luminosity suddenly changes from one (roughly) constant
value to another.  In this case, a natural description of the
properties of changes in contrast would be {\it to count} the number
of such changes that appear along a segment of size $r$, oriented in a
given direction. The statistical distribution of the position and size
of these jumps in contrast would give relevant information about an
ensemble of such images.  However, in real images this counting is not
enough to describe well the properties of contrast changes.  This is
because these changes do not follow such a simple pattern: some of
them can be sharp (the notion of sharp changes has to be given), but
there will also be all types of softer textures that would be lost by
just counting the most noticeable changes. It is then necessary to
consider all of them as a whole.  A natural way to deal with contrast
changes is by defining a quantity that accumulates all of them,
whatever their strength, contained inside a scale $r$, that is:

\eq
\epsilon_r(\vec{x})  \:=\: 
 \frac{1}{r^2} 
\int_{x_1 - \frac{r}{2}}^{x_1 + \frac{r}{2}} dx_1'
\int_{x_2 - \frac{r}{2}}^{x_2 + \frac{r}{2}} dx_2'  \;\;  
\mid \nabla C({\vec{x}'}) \mid   \;\;   .
\label{eq:square}
\en

\noindent
Hereafter, the contrast $C(\vec{x})$ is taken as  
$C(\vec{x})=I(\vec{x})-\langle I \rangle$, where $I(\vec{x})$ is the 
field of luminosities and $\langle I \rangle$
its average value across the ensemble. The bidimensional integral 
\footnote{
The variables $x_1^{\prime}$, $x_2^{\prime}$ are the
components of the vector $\vec{x}^{\prime}$ and $\mid \mid$ denote 
the modulus of a vector.} 
on the right hand side,
defined on the set of pixels contained in a square of linear size $r$,
is a {\it measure} of that square\footnote{
 Any measure of a set has
the property of being \emph{additive}, that is: if one splits the set
in pieces, the measure of the set is equivalent to the sum of the
measures of all its pieces.}
. It is divided by the factor $r^2$, which is the
usual Lebesgue measure (which we will denote $\lambda$) of a square of
linear size $r$. The quantity $\epsilon_r(\vec{x})$ can then be regarded
as the ratio between these two measures. More generally,
we can define the measure $\mu $ of a subset $A$ of image pixels as

\eq
\mu (A) \:=\: \int_{A} d\vec{x}' \; \mid \nabla C(\vec{x}') \mid  \;\; ,
\label{eq:measure}
\en

\noindent
($\int_{A} d\vec{x}'$ means bidimensional integration over the set $A$)
what we will call the \emph{Edge Measure} (EM) of the subset $A$. 
It is also possible to generalize the definition of $\epsilon_r$
for any subset $A$:

\eq
\epsilon_A\: =\: \frac{\mu (A)}{\lambda(A)} \hspace*{1cm}.
\label{eq:edge_content}
\en

\noindent
$\epsilon_A$ is a density, a quantity that compares
how much the distribution of $|\nabla C|$ deviates from being
homogeneously distributed on $A$. We shall call it the \emph{Edge 
Content} (EC) of the set $A$. Denoting $B_r(\vec{x})$ the ``ball''\footnote{
``Ball'' in a mathematical sense: it does not need to be true circles, but
could also be squares or diamonds. In fact, along this paper the 
``balls'' will be taken as squares.}
of radius $r$ centered around $\vec{x}$, it is clear that
$\epsilon_r(\vec{x})\equiv \epsilon_{B_r(\vec{x})}$, so this definition
generalizes the previous one. We shall refer to $\epsilon_r(\vec{x})$
as the Edge Content of $\vec{x}$ at the scale
$r$. Notice that $\epsilon_{dr}(\vec{x})$ is the
$\mu $-density at $\vec{x}$, $|\nabla C|(\vec{x})$. Thus,
eq.~(\ref{eq:measure}) can be symbolically expressed as 

\eq
d\mu (\vec{x}) = |\nabla C|(\vec{x}) d\vec{x} \;\; . 
\en

\noindent
The main point is that contrast changes are distributed over the image
in such a way that the Edge Content has rather large contributions even from
pixels that are very close together.  As a consequence, the measure
presents a rather irregular behavior. But it is precisely in its
irregularity where the information about the contrast lies. Let us
assume for instance that, as $r$ becomes very small, the measure
behaves at every point $\vec{x}$ as

\eq
\mu \left(B_r(\vec{x})\right)\:\equiv\:\mu _r(\vec{x}) \:=\: \alpha(\vec{x}) \;
r^{h(\vec{x})+d} \;\; ,
\label{eq:exponent}
\en

\noindent
where $d=2$ is the dimension of the images. For convience a shift in the
definition of the exponent of $\mu _r(\vec{x})$ has been introduced. The 
Edge Content then verifies~:

\eq
\epsilon_r(\vec{x})\: =\: \alpha(\vec{x})\; r^{h(\vec{x})}\;\; ,
\label{eq:exponent_epsilon}
\en

\noindent
and this choice of $h(\vec{x})$ removes trivial dependencies on $r$. 
The exponent $h(\vec{x})$ is just the singularity exponent of $|\nabla 
C|(\vec{x})$. That is, $h<0$ indicates a divergence of $|\nabla 
C|(\vec{x})$ to infinity, while $h>0$ indicates finiteness and 
continuous behavior. The greater the exponent the smoother is the 
density around that point. In general, we will talk about ``singularity'' 
even for the positive $h$ cases.

\indent 
Thus, the meaning of eq.~(\ref{eq:exponent}) is that all the
points are singular (in this wide sense), and that the singularity
exponent is not uniform. This irregular behavior would allow us to 
classify the pixels in a given image: the set of pixels with a singularity 
exponent contained in the interval $[h -\Delta_h, h + \Delta_h]$ (where 
$\Delta_h$ is a small positive number) define a class $F_h$.  These 
classes are the \emph{fractal components} of the image.  The smaller the 
exponent the more singular is the class, and the most singular component 
is the one with the smallest value of $h$. A measure $\mu $ verifying 
eq.~(\ref{eq:exponent}) in every point $\vec{x}$ is called a 
\emph{multifractal measure}. 

\indent 
We will refer to these components of the images as fractal sets
because it will be seen that they are of a very irregular nature.  One
way to characterize the odd arrangement of the points in fractal sets
is by just counting the number of points contained inside a given
ball of radius $r$; we will denote by $N_r(h,\Delta_h)$ this number
for the set $F_h$. As $r \rightarrow 0$ it is verified that

\eq
N_r(h, \Delta_h) \:\sim\: r^{D(h)}.
\label{eq:component_num}
\en

\noindent
This new exponent, $D(h)$, quantifies the size of the set of pixels
with singularity $h$ as the image is covered with small balls of
radius $r$. It is the fractal dimension of the associated fractal
component $F_h$, and the function $D(h)$ is called the
\emph{dimension spectrum} (or the \emph{singularity spectrum}) of the
multifractal. It is also worth estimating the probability for a ball
of radius $r$ to contain a pixel belonging to a given fractal
component.  Its behavior for small scales $r$ is:

\eq
Prob(h, \Delta h)  \:\sim\: r^{d-D(h)}
\;\; ,
\label{eq:component_prob}
\en

\noindent
where $d=2$ as before. Notice that, since $r$ is small, this quantity
increases as $D(h)$ approaches $d$.\footnote{
It can be proved that any set contained in a $d$-dimensional
real space has fractal dimension $D$ smaller or equal to $d$.
}

\indent
The multifractal behavior also implies a \emph{multiscaling} 
effect: as every fractal component possesses its own fractal dimension, 
each of them changes differently under changes in the scale. Thus, 
although every component has no definite scale and is invariant under the
scale transformations, scale invariance is broken for a given 
image: the points in the image transform with different
exponents, according to the fractal component to which they
belong. This leads to the presence of intermittency effects: if 
the image were enlarged, its statistical properties would
change. Self-similarity is lost.  However, scale invariance is
restored in the averages over an ensemble of such images.

\indent 
At this point one may wonder if a multifractal behavior is actually
realized in natural images.  One way to check its validity is to 
verify that eq.~(\ref{eq:exponent}) is indeed correct.  This in turn
leads us to the technical problem of how to obtain in practice the
exponent $h(\vec{x})$ at a given pixel $\vec{x}$ of the image.  This
is a local analysis where the singularity is found by looking at the
neighbourhood of the point $\vec{x}$. It has the advantage that it not
only provides us with a tool to check the multifractal character of
the measure but it also gives a way to decompose each image in 
its fractal components.  These issues are addressed in the next
Section.  The multifractality of the measure can also be studied by
means of statistical methods.  In this case one looks for the effect
of the singulaties on the moments of marginal distributions of, e.g.,
the Edge Content defined in eq.~(\ref{eq:square}). This is done in Section
\ref{section:statistics}, where the important notions of
self-similarity, extended self-similarity and multiplicative processes
are explained.  Once the relevant mathematical tools are given, these
are used in Section \ref{section:results} to analyze natural images.
First in subsection \ref{subsection:exp.wavelet} we check that the
measure introduced in eq.~(\ref{eq:measure}) is indeed multifractal.
Examples of fractal components of natural images are also presented in
the same section.  After this we study the consistency between the
singularity analysis and the statistical analysis of the measure
(subsection \ref{subsection:exp.SS}).  The results of the numerical
study of self-similarity properties of the contrast $C(\vec{x})$ and
of the measure density $|\nabla C|(\vec{x})$ are presented in
subsection \ref{subsection:exp.multiaffine}.  The discussion of the
results and perspectives for future work are given in the last
section.  Some technical aspects are included in the Appendix.

\section{Singularity analysis: The wavelet transform}
\label{section:wavelet}

\indent

In this section we present the appropriate techniques to deal with 
multifractal measures, and especially with those adapted to natural images.
To check if $\mu $ defines a multifractal measure and to obtain
the local exponent $h(\vec{x})$, we need to analyze the dependence of
$\mu (B_r(\vec{x}))$ on the scale $r$.  Unfortunately, direct
logarithmic regression performed on eq. (\ref{eq:exponent}) yields
rather coarse results when it is applied on discretized images,
allowing only the detection of very sharp, well isolated features. It
is then necessary to use more sophisticated methods. A convenient
technique for this purpose \cite{Mall+92,Arn+95,Arn96} is the
\emph{wavelet transform} (see e.g.  \cite{Dau92}).  Roughly speaking,
the
wavelet transform provides a way to interpolate the behavior of the
measure $\mu $ over balls of radius $r$ when $r$ is a non-integer
number of pixels. Besides, if the image is a discretization of a
continuous signal, the wavelet analysis is the appropriate
technique to retrieve the correct exponents, as it is explained
later on.

\indent
The wavelet transform of a measure $\mu $ is a function that is not 
localized, neither in space nor in frequency. Its arguments are then the 
site $\vec{x}$ and the scale $r$. It also involves an appropriate,
smooth interpolating function $\Psi(\vec{x})$. More precisely, the wavelet
transform of $\mu $ is:

\begin{equation}
T^r_{\Psi}d\mu (\vec{x}) \; \equiv\; \int d\mu (\vec{x}')\: 
\Psi_{r} (\vec{x}-\vec{x}') \; =\; \frac{d\mu }{d\vec{x}}\otimes \Psi_r
(\vec{x})\;\;
,
\label{eq.def_wavelet_transform}
\end{equation}

\noindent
where $\Psi_r (\vec{x})\equiv \frac{1}{r^d} \Psi(\frac{\vec{x}}{r})$.
The important point here is that for the case of a positive 
measure it can be proved that:

\eq
T^r_{\Psi} d\mu (\vec{x}_0) \sim r^{h(\vec{x}_0)}
\label{eq:exponent_wv}
\en

\noindent
if and only if $\mu $ verifies eq.(\ref{eq:exponent}) ($h$ being exactly 
the same) and $\Psi$ decreases fast enough (see \cite{Arn96} and 
references therein). In that way, this property allows us to extract the 
singularities directly from $T^r_{\Psi}d\mu $. It is very important to 
remark that $\Psi$ can be a positive function, what makes an 
essential difference with respect to the 
analysis of multiaffine functions (see section 
\ref{subsection:multiaffine}).\footnote{
This means that the exponents can be obtained with a
\emph{non-admissible} wavelet $\Psi$, i.e. a function with a non-zero
mean}
%
%

\indent
In general the ideal, continuous measure $d\mu (\vec{x})$ is unknown. 
In these cases the data is given by a discretized sampling, which in 
our case is a collection of pixels.  One can reasonably argue that 
the pixel is the result of the convolution of an ideal signal with 
a compact support function $\chi_a(\vec{x})$ (describing, e.g., the resolution 
$\sim a$ of the optical device). In this case, under convenient hypothesis, the 
wavelet transform yields the correct projection of the ideal signal.
Let us call $\vec{x}_{(n_1,n_2)}\equiv\vec{x}_{\vec{n}}$ the positions 
of the pixels in 
the discretized sample and $C_{\vec{n}}$ the values of the discretized
version of the contrast; then

\eq
C_{\vec{n}}\; =\; T^a_{\chi} C(\vec{x}_n) \,\equiv\, C\otimes \chi_a
(\vec{x}_{\vec{n}}) \; ,
\label{eq:discretized}
\en

\noindent
Taking the sample $\{ \vec{x}_{\vec{n}} \}$ in such a
way that the centers of adjacent pixels are at a distance 
larger than $a$, it is verified that:

\eq
\nabla C_{\vec{n}}\; \equiv\; \left( C_{(n_1+1,n_2)}-C_{(n_1,n2)}\: ,\: 
C_{(n_1,n_2+1)}-C_{(n_1,n2)} \right)\; \sim\; \nabla C\otimes \chi_a
(\vec{x}_{\vec{n}})
\en

\noindent
and so the discretized version $\mu ^{(dis)}$ of the ideal measure $\mu $, can be
considered as a wavelet coefficient of the continuous $\mu $ with the
wavelet $\chi$ at the scale $a$, namely

\eq
d\mu ^{(dis)}_{\vec{n}}\sim d\vec{x}\;\;|\nabla C|\otimes \chi_a
(\vec{x}_{\vec{n}}) \;=\;d\vec{x}\;\; T^a_{\chi}d\mu 
(\vec{x}_{\vec{n}}) \;\; .
\en

\noindent
Taking into account that the convolution is associative, the discretized
wavelet analysis of $d\mu _{\vec{n}}^{(dis)}$ with a wavelet $\Psi$ can be 
simply expressed as:

\eq
T^r_{\Psi} d\mu ^{(dis)} (\vec{x}_{\vec{n}})\sim  
T^r_{\Psi\otimes \chi_{\frac{a}{r}}} d\mu 
(\vec{x}_{\vec{n}}) 
\en

\noindent
that is, analyzing the discretized measure $\mu ^{(dis)}$ with a wavelet $\Psi$
is equivalent to analyze the continuous measure $\mu $ with a wavelet 
$\Psi\otimes\chi_{\frac{a}{r}}$. But if the scale $r$ is large enough
compared to the photoreceptor extent $a$, $\chi_{\frac{a}{r}}(\vec{x})\sim
\delta (\vec{x})$, which does not depend on $r$. Thus, 
if the internal size $a$ is small enough to allow
fine detection, we can recover the
correct exponent at the point $\vec{x}_{\vec{n}}$ by just analyzing the
discretized measure. Conversely, the extent $a$ imposes a lower cut-off in the
details that can be observed in any image constructed by pixels: as
it is just the size of the pixel, 
the estimation of the singularity exponent at a given point requires 
to perform the analysis over scales containing several pixels.

\indent
This analysis is similar to that performed in   
\cite{Mall+91,Mall+92}: any image is analyzed by means of its scaling 
properties under wavelet projection. There are however important differences 
between the study performed by these authors and the present work.
The first difference concerns the basic scalar field to be analyzed:
Mallat and Zhong considered $C(\vec{x})$ instead of $|\nabla C|(\vec{x})$, 
which we prefer because, as it will be shown later on,  it is the $\mu $-density of a 
multifractal measure $\mu $ (see Section \ref{section:results}). Another 
difference is methodological: in this paper we intend to classify the 
points according its singularity exponent to form every fractal 
component, while the previous papers were devoted to obtain only the 
wavelet transform modulus maxima, a concept related but different from what we 
will call the most singular component, that is, only one (although the 
most important) of the fractal components. The third difference concerns 
motivation: we want to classify every point with respect to a 
hierarchical scheme, while these authors search a wavelet-based optimal 
codification algorithm, in the sense of providing an (almost) perfect 
reconstruction algorithm from the wavelet maxima.

\section{Statistical self-similarity of natural images}
\label{section:statistics}

\indent
In this section we introduce the basic concepts of the statistical
approach, defining at the same time new variables which are closer to
the wavelet analysis.

\subsection{SS and ESS}
\label{subsection:SS-ESS}

\indent

 From the statistical point of view, $\mu _r$ is a random variable, in the 
sense that given a random point $\vec{x}$ in an arbitrary image 
$\cal{I}$ the value of $\mu ^{\cal{I}}_r(\vec{x})$ cannot be 
deterministically predicted. 


\indent
By definition, the multifractal structure is a multiscaling 
effect. This multiscaling should be somehow apparent from the dependence 
of the probability density of $\mu _r$ on the scale parameter $r$. 
To determine the whole probability density function of $\mu _r$ requires a 
large dataset, especially for the rare events; besides its dependence on $r$
could mix multiscaling effects in a complicated way. On the 
contrary, the $p$-moments of $\mu _r$ (that is, the expectation values of
$\mu _r$ raised to the $p$-th power) suffice to characterize completely
the probability density,\footnote{
Provided they do not diverge too fast 
with $p$. }
and given that they are averages of powers of $\mu _r$ they are  
likely to depend on $r$ in a simple way, somehow related to the
behavior of the measure in eq.~(\ref{eq:exponent}). 

\indent
It is more convenient to deal with the $p$-moments of normalized
versions of $\mu _r$ such as the Edge Content $\epsilon_r\equiv \frac{1}{r^2}
\mu _r$ or the wavelet projections $|T^r_{\Psi}d\mu |$. They are
normalized in such a way that the first order moments, $\langle
\epsilon_r\rangle$ and $\langle |T^r_{\Psi}d\mu | \rangle$, do not depend
on $r$: the trivial factor $r^{p d}$ is hence removed from the random
variable. 

\indent
As we will see later on, the moments $\langle
\epsilon_r^p \rangle$ and $\langle |T^r_{\Psi}d\mu |^p\rangle$  
exhibit the remarkable scaling properties of Self-Similarity
(SS):

\eq
\langle \epsilon_r^p \rangle \; =\; \alpha_p\: r^{\tau_p}
\label{eq:SS}
\en

\noindent
(analogously for $|T^r_{\Psi}d\mu |$ with exponents $\tau_p^{\Psi}$) and of Extended Self-Similarity
(ESS) (referred to the moment of order $2$) \cite{Ben+93b,Ben+93c,Ben+95}:

\eq
\langle \epsilon_r^p \rangle \; =\; A(p,2)
\left[ \langle \epsilon_r^2 
\rangle\right]^{\rho(p,2)}\;\;,
\label{eq:ESS}
\en

\noindent
(and similarly for $|T^r_{\Psi}d\mu |$ with exponents
$\rho^{\Psi}(p,2)$).  ESS has been referred to the moment of order
$2$, but eq~(\ref{eq:ESS}) trivially implies that any moment can be
expressed as a power of the moment of order $q$ with an exponent
$\rho(p,q)=\frac{\rho(p,2)}{\rho(q,2)}$.

\indent
In this way, we have the sets of SS exponents $\tau_p$ and ESS
exponents $\rho(p,2)$ for $\epsilon_r$ and similarly the exponents
$\tau_p^{\Psi}$ and $\rho^{\Psi}(p,2)$ for $T^r_{\Psi}d\mu $. Notice that
if SS holds also does ESS, in which case there is a simple relation
between $\tau_p$ and $\rho(p,2)$:

\eq
\tau_p\; =\; \tau_2 \: \rho(p,2) \; \; .
\label{eq:rel_SS-ESS}
\en

\indent 
The actual dependence of $\tau_p$ and $\rho(p,2)$ on $p$
determines the fractal structure of the system. A trivial dependence
$\tau_p\propto p$ would reveal a monofractal structure. 
A different dependence on $p$ indicates the existence of 
a more complicated (multifractal) geometrical structure.

\subsection{The multiplicative process}
\label{subsection:Log-Poisson}

\indent
A random variable subtending an area of linear size $r$ and which
possesses SS (eq.~(\ref{eq:SS})) and ESS (eq.~(\ref{eq:ESS})) can be
described statistically by means of a multiplicative process
\cite{Ben+93a,Nov94}.  The statistical formulation is rather simple:
given two different scales $r$ and $L$ such that $r<L$, there is a
simple stochastic relation between the Edge Contents at these two scales~:

\eq
\epsilon_r\; \doteq \; \alpha_{rL} \: \epsilon_L
\en

\noindent
where $\alpha_{rL}$ is a random variable, independent of the Edge Content, 
which describes how the change between these two scales takes place. The
symbol ``$\doteq$'' indicates that both sides of the equation are
distributed in the same way, but this does not necessarily imply the
validity of the relation at every pixel of a given image. The random
variables $\alpha_{rL}$ between all possible pairs of scales $r$ and
$L$ are said to define a \emph{multiplicative process}.

\indent
Let us understand in a more intuitive way the meaning of this process. It is
rather clear that given the
distribution of $\epsilon_L$ at a fixed large scale $L$, to compute
the distribution of the Edge Content $\epsilon_r$ it is enough to know the
distribution of $\alpha_{rL}$. But let us now introduce an
intermediate scale $r'$. One could also obtain first $\epsilon_{r'}$
using $\alpha_{r'L}$ and then $\epsilon_{r}$ by means of
$\alpha_{rr'}$. Thus, the multiplicative process must verify the
cascade relation:

\eq
\alpha_{rL}\; =\; \alpha_{rr'}\:\alpha_{r'L} \;\; ,
\label{eq:cascade}
\en

\noindent
that tells us that the variables $\alpha_{rL}$ are \emph{infinitely
divisible}.  Infinitely divisible stochastic variables are then
completely characterized by their behavior under infinitely small
changes in the scale, that is, by $\alpha_{r,r+dr}$. The simplest
example of an infinitely divisible random variable $\alpha_{rL}$ is the
one having a binomial infinitesimal distribution, namely:

\eq
\alpha_{r,r+dr}\; =\;
\left\{\begin{array}{ccc}
\\
1-\Delta \frac{dr}{r} & , & \mbox{with probability $1-[d-D_{\infty}] 
\frac{dr}{r}$}\\ \\
\beta (1-\Delta \frac{dr}{r}) & , & \mbox{with probability $[d-D_{\infty}] 
\frac{dr}{r}$} \\ \\
\end{array} \right.
\label{eq:Log-Poisson_inf}
\en

\noindent
The form of this process is determined by its compatibility with
eq.~(\ref{eq:cascade}); see for instance
\cite{She94,She+95,Cas96} for a detailed study. Moreover,
there are only two free parameters: it is verified that $d-D_{\infty}=
\Delta/(1-\beta)$. They have also to satisfy the following bounds:
$0<\beta<1$ and $0<\Delta <1$.

\indent
The infinitesimal process (\ref{eq:Log-Poisson_inf}) can be easily
interpreted in terms of the Edge Content and the measure density:

\begin{itemize}

\item
{\bf Probability $ \; = 1-[d-D_{\infty}] \frac{dr}{r}$ :} this is the most
likely situation.  In this case the Edge Content (and consequently the measure)
changes smoothly under small changes in scale: $\alpha_{r,r+dr}$ is
close to one and its variation is proportional to $d \ln r$.

\item
{\bf Probability $ \; = [d-D_{\infty}] \frac{dr}{r}$ :} in this unlikely case,
under infinitesimal changes in scale the Edge Content undergoes finite variations, 
which is reflected in the fact that $\alpha_{r,r+dr}$ deviates from
one by 
a factor $\beta$. The $\mu $-density $|\nabla C|$ must then be divergent 
somewhere along the boundary of the ball of radius $r$, and the parameter 
$\beta$ is a measure of how sharp this divergence is.

\end{itemize}

\indent
The process for non-infinitesimal changes in scale can be easily
derived from the infinitesimal process presented in
eq~(\ref{eq:Log-Poisson_inf}).  In this case the random variable
$\alpha_{rL}$ follows a Log-Poisson process \cite{She+95}; its 
probability distribution $\rho_{\alpha_{rL}}(\alpha_{rL})$ has the
form:

\eq
\rho_{\alpha_{rL}}(\alpha_{rL})\; =\; \left[\frac{r}{L}\right]^{d-D_{\infty}}
\: \sum_{n=0}^{\infty} \frac{(d-D_{\infty})^n}{n!}\: 
\left[\ln\frac{L}{r}\right]^n\: \delta (\alpha_{rL} -\beta^n\:\left( 
\frac{r}{L} \right)^{-\Delta})
\label{eq:Log-Poisson}
\en

\indent
If a multiplicative process is realised the Edge Content has the SS and the ESS
properties, eqs.~(\ref{eq:SS}) and (\ref{eq:ESS}). Thus, knowledge of
the ESS exponents $\rho(p,2)$ and of $\tau_2$ is enough to compute
$\tau_p$ using eq.~(\ref{eq:rel_SS-ESS}). For the Log-Poisson process
the ESS exponents are:

\eq
\rho(p,2) \; =\; \frac{p}{1-\beta}\: -\: \frac{1-\beta^p}{(1-\beta)^2}
\label{eq:rho_S-L}
\en

\noindent
which depend only on $\beta$. This process was proposed in
\cite{She94} for turbulent flows, so we will refer to it 
either as the She-Leveque (S-L) or as the Log-Poisson
model. The SS exponents are:

\eq
\tau_p\; =\; -\Delta p \: +\: (d-D_{\infty}) (1-\beta^p)
\label{eq:tau_S-L}
\en

\noindent
which compared with eq.~(\ref{eq:rel_SS-ESS}) gives the following set of 
relations~:

\eq
\left\{
\begin{array}{ccccc}
\Delta & = & -\frac{\tau_2}{1-\beta} & & \\
d-D_{\infty} & = & -\frac{\tau_2}{(1-\beta)^2} & = & \frac{\Delta}{1-\beta} 
\end{array}
\right.
\en

\noindent
This means that the model has only two free parameters, that can be
chosen to be $\tau_2$ and $\beta$ or $\Delta$ and $D_{\infty}$, for
instance.  The geometrical interpretation of the last two is very
interesting \cite{Tur+98}, and will be explained in the next subsection.
For the time being let us notice that the maximum value of the
EC, $\parallel \epsilon_r \parallel_{\infty}$, also follows a power
law with exponent $-\Delta$,

\eq
\parallel \epsilon_r \parallel_{\infty}\; =\; \alpha_{\infty}\: r^{-\Delta}
\en

\noindent
and thus the parameter $\Delta$ characterizes the most divergent 
behavior present in natural images.

\subsection{SS and multifractality}
\label{subsection:ssandmulti}

\indent
Let us now analyze in more detail the geometrical meaning of the model
and its relation with the singularities described in
Section~\ref{section:wavelet}.  For a multifractal measure $\mu $ with
singularity spectrum $D(h)$ there is an important relation between the
SS exponents $\tau_p$ and $D(h)$.  Let us denote by $\rho_h(h)$ the
probability density of the distribution of the singularity exponents
$h$ in the image. Then, partitioning the image pixels according to
their values of $h$, eq.~(\ref{eq:exponent_epsilon}) can be expressed,
for $r$ small enough, as:

\eq
\langle \epsilon_r^p \rangle \; \sim\; \int dh \: \rho_h(h) \langle 
\alpha^p\rangle_{_{F_h}} \: r^{hp +d-D(h)} \;\; .
\label{eq:preSP}
\en

\noindent
The factor $r^{d-D(h)}$ comes from the probability of a randomly
chosen pixel being in the fractal $F_h$ (see
eq.~(\ref{eq:component_prob})) and guarantees the correct
normalization of $\langle\alpha^p\rangle_{_{F_h}}$ (see
e.g. \cite{Fri95,Arn96} for details).  For very small $r$ the
application of the saddle point method to eq.~(\ref{eq:preSP}) yields:

\eq
\langle \epsilon_r^p \rangle \; =\; \alpha_p r^{\tau_p}\;  \propto\; 
r^{\minimo{h}{ph + d- D(h)}} 
\en

\noindent
This gives a very interesting relation between $\tau_p$ and $D(h)$: 
$\tau_p$ is the Legendre transform of $D(h)$. The important point is that 
$D(h)$ is easily expressed in terms of $\tau_p$ by means of another 
Legendre transform, namely:

\eq
D(h)\; =\; \minimo{p}{ph +d -\tau_p}\;\; .
\label{eq:sing_spectrum}
\en

\noindent
If one has  a model for the $\tau_p$'s (the S-L model, 
eq.~(\ref{eq:tau_S-L})) the dimension spectrum can be predicted:

\eq
D(h)\; = \; D_{\infty} - \frac{h+\Delta}{\ln\beta}
\left[1-\ln\left( -\frac{h+\Delta}{(d-D_{\infty})\ln\beta}\right) \right]\;\; .
\label{eq:Dh_S-L}
\en

\begin{figure}[htb]
\begin{center}
\hspace*{0cm}
\epsfxsize=14cm
\epsfbox[50 50 410 302]{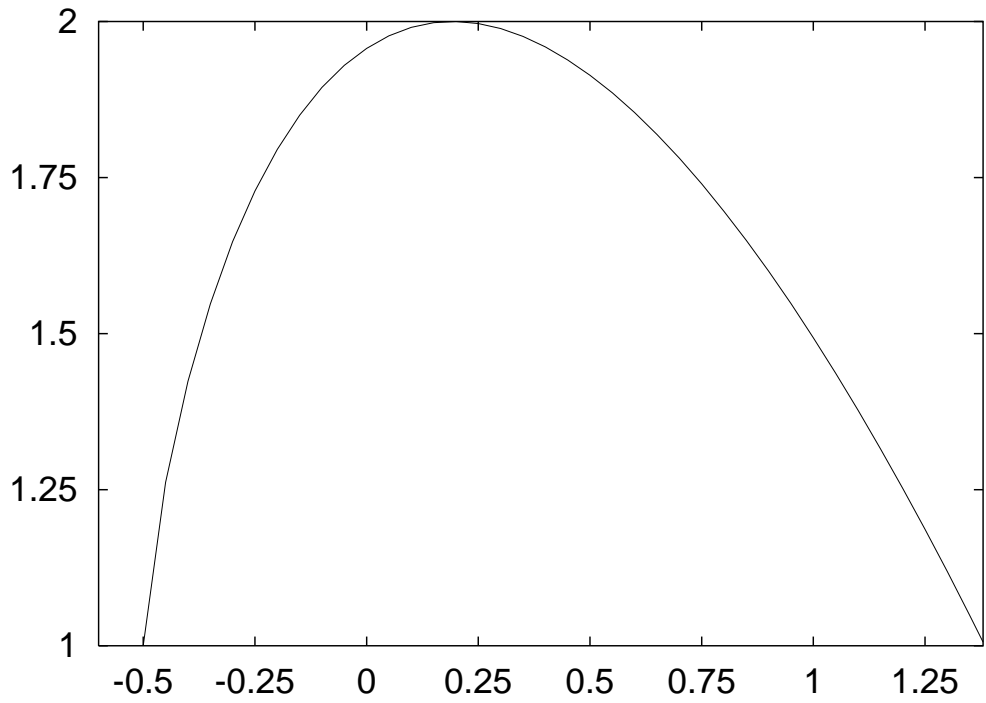}
\caption{
Singularity spectrum of the Log-Poisson model for $\beta=0.5$ and 
$\tau_2=-0.25$. For these values $D_{\infty}=1.$ and $\Delta=0.5$
(see text)
}
\label{fig:Dh_S-L}
\end{center}
\end{figure}

\noindent
This is represented in Fig.~\ref{fig:Dh_S-L}. There are natural cut-offs
for the dimension spectrum; in particular, there cannot be any exponent
$h$ below the minimal value $h_{\infty}=-\Delta$. Thus, $\Delta$ defines the
most singular of the fractal components, which has fractal dimension
$D(h_{\infty})=D_{\infty}$.  We could use these properties of the
most singular fractal (its dimension and its associated exponent) to
characterize the whole dimension spectrum, which is in agreement with the
arguments given in section~\ref{subsection:Log-Poisson}.\footnote{
Remember that $\beta$ can be expressed in terms of $\Delta$ and
$D_{\infty}$:  $\beta=1-\Delta/(d-D_{\infty})$ 
}

\subsection{Multiaffinity}
\label{subsection:multiaffine}

\indent

Related to multifractality there exists the simpler but
more unstable property of multiaffinity \cite{Ben+93a}. This is a 
characterization of chaotic, irregular scalar functions that somewhat 
generalizes the concepts of continuity and differentiability.

\indent
First, let us give the concept of \emph{H{\"o}lder exponent}: A scalar 
function $F(\vec{x})$ is said to be H{\"o}lder of exponent $h_F(\vec{x}_0)$ 
at a given point $\vec{x}_0$ if for any point $\vec{y}$ close enough to 
$\vec{x}$ the following inequality holds:

\eq
|F(\vec{y})-F(\vec{x}_0)|\: <\: A_0 
|\vec{y}-\vec{x}_0|^{h_F(\vec{x}_0)}\;\; , \label{eq:Holder}
\en

\noindent
$A_0$ being a constant depending on the point $\vec{x}_0$. We define the 
H{\"o}lder exponent of $F(\vec{x})$ at $\vec{x}_0$ as the maximum of the 
exponents $h_F(\vec{x}_0)$ verifying eq.~(\ref{eq:Holder}). Defining 
the Linear Increment (LI) of $F(\vec{x})$ by a displacement vector 
$\vec{r}$ as $\delta_{\vec{r}} F(\vec{x})\equiv 
|F(\vec{x}+\vec{r})-F(\vec{x})|$, if the function $F$ has a H{\"o}lder
exponent $h_F(\vec{x})$ at $\vec{x}$ then:

\eq
\delta_{\vec{r}} F (\vec{x})\; \sim\; \alpha_F(\vec{x}) r^{h_F(\vec{x})}
\label{eq:multiaffine}
\en

\noindent
which is similar to eq.~(\ref{eq:exponent_epsilon}). In the same spirit,
we say that a function is \emph{multiaffine} if for every point $\vec{x}$
eq.~(\ref{eq:multiaffine}) is verified.

\indent
Multiaffinity is a more intuitive concept than multifractality, because
the H{\"o}lder exponents are good characterizations of Taylor-like local
expansions of the function $F(\vec{x})$. For instance, $h_F(\vec{x})>0$
means that the function is continuous at $\vec{x}$, $h_F(\vec{x})>1$ implies
that the function has a continuous first derivative and so on. It also
works in the other sense: a function having exponent $h_F=-1$ behaves as
``bad'' as a $\delta$-function (See the Appendix for a brief tutorial 
about H{\"o}lder exponents of simple functions). Even more: a function $F$ has 
H{\"o}lder exponent $h_F$ at a given point if and only if its first order 
derivatives have H{\"o}lder exponent $h_F-1$ at the same point. 

\indent
Besides, multiaffinity implies multifractality in the following sense:  
given a measure $\mu $, if its density $\frac{d\mu }{d\vec{x}}$ is a
multiaffine function then $\mu $ is a multifractal measure having exactly
the same exponents as $\frac{d\mu }{d\vec{x}}$ at every point.\footnote{ 
Notice that there is a shift of size $d$ in the
definition of the exponents of $\mu $ (eq.~(\ref{eq:exponent})) which is
not present in the definition of the H{\"o}lder exponents of a multiaffine 
function (eq.~(\ref{eq:Holder})). 
}
For natural images and taking $F=|\nabla C|$, this fact can be expressed
as:

\eq
\epsilon_r \;\sim\; T^r_{\Psi} d\mu \;\sim\; \delta_{\vec{r}} |\nabla C|
\label{eq:statistical}
\en

\noindent
in the sense that the three variables have the same statistical
dependence on $r$. This relation can be also expressed in terms of the SS
exponents.  Let us denote by $\tau_p$ the SS exponents of
$\epsilon_r$, by $\tau_p^{\Psi}$ those of $T^r_{\Psi} d\mu $ and by
$\tau_p^{\nabla C}$ those of $\delta_{\vec{r}} |\nabla C|$. Then the
statistical relation in  eq.~(\ref{eq:statistical}) implies that:

\eq
\tau_p \; =\; \tau_p^{\Psi}\; =\; \tau_p^{\nabla C}
\en

\noindent
We are also interested in the relation between $\tau_p$ and the SS exponents
$\tau_p^C$ of the moments of $\delta_{\vec{r}}C$. 
If $C$ exhibits the multiaffine behavior shown in eq.~(\ref{eq:multiaffine}),
recalling the meaning in terms of differentiability of the H{\"o}lder
exponents, we have that $\delta_{\vec{r}} |\nabla C|(\vec{x})$ verifies:

\eq
\delta_{\vec{r}} |\nabla C| (\vec{x})\; \sim\; \alpha_{_{|\nabla 
C|}}(\vec{x}) \; r^{h_C(\vec{x})-1} 
\label{eq:shift-1}
\en

\noindent
This can be statistically interpreted as:

\eq
\epsilon_r\; \sim  \;\delta_{\vec{r}} |\nabla C|\; \sim\; \frac{1}{r}\:  
\delta_{\vec{r}} C 
\label{eq:relation_LIs}
\en

\noindent
which reflects the shift by -1 in the exponents of the derivative of 
$C$.\footnote{
Let us note that the relation $r \epsilon_r\; \sim  \; \delta_{\vec{r}} C$ 
is the analog of the Kolmogorov hypothesis of local similarity 
\cite{Arn96}
}
In terms of the SS exponents, this implies that:

\eq
\tau_p^C\; =\;  \tau_p^{\nabla C} + p \; = \; \tau_p +p
\label{eq:relation_LIs_taup}
\en

\noindent
Let us remark that the singularity spectra associated to the 
variables $\delta_{\vec{r}} C$, $\delta_{\vec{r}} |\nabla C|$ and 
$T^r_{\Psi} d\mu $ are the Legendre transforms of the SS exponents   
$\tau_p^C$, $\tau_p^{\nabla C}$ and $\tau_p^{\Psi}$, respectively 
(a fact that can be derived in the same way as 
eq.~(\ref{eq:sing_spectrum})).

\indent
There still remains the converse question: if multifractality implies 
multiaffinity. Unfortunately, this is not true (see \cite{Dau92}). 
In fact, the $|F(\vec{y})-F(\vec{x}_0)|$ term in eq.~(\ref{eq:Holder})
suggests the existence of a pseudo-Taylor expansion, in the sense that:

\eq
F(\vec{y})\;\approx\; p_n(\vec{y}-\vec{x}_0)\: +\:
A_0\: |\vec{y}-\vec{x}_0|^{h_F(\vec{x}_0)}
\label{eq:pseudoTaylor}
\en

\noindent
where $p_n(\vec{y}-\vec{x}_0)$ is a polynomial of degree $n$ in the
projection of the vector $\vec{y}-\vec{x}_0$ along its direction. We
should thus generalize the concept of H{\"o}lder exponent: given a point
$\vec{x}_0$, we will say that this point possesses a H{\"o}lder exponent
$n\: < \: h_F(\vec{x}_0)\: <\: n+1$ ($n$ integer) if there exists a 
polynomial $p_n(\vec{x})$ of degree $n$ and a constant $A_0$ such that

\eq
|F(\vec{y})-p_n(\vec{y}-\vec{x}_0)|\; <\; A_0\: 
|\vec{y}-\vec{x}_0|^{h_F(\vec{x}_0)} 
\label{eq:Holder_general}
\en

\noindent
and $h_F(\vec{x}_0)$ is the maximum of the exponents verifying such relation.

\indent 
Multifractal measures with no multiaffine densities (as defined in
eq. (\ref{eq:Holder})) have been studied in
\cite{Bac+93,Arn96}. The problem can be explained by the masking
presence of the regular, polynomial part which stands for global,
non-localized effects in the signal field $F(\vec{y})$. To remove this part, those
authors propose a generalization of the Linear Increment. Instead of Linear
Increments of the function, wavelet transforms of the appropriate
one-dimensional restriction are considered, taking the scalar field as
a real-valued measure density. This wavelet
transform of the function $F(\vec{x})$ is defined as

\eq
T^{\vec{r}}_{\Psi} F (\vec{x}_0)\;\equiv\; \int ds \; 
F(\vec{x}_0+s\frac{\vec{r}}{r}) \Psi_r(s)  \; ,
\label{eq:1d_restriction}
\en

\noindent
where $\Psi$ is a real, one-dimensional analyzing wavelet.  Since the
integral is one-dimensional the wavelet at the scale $r$ is
$\Psi_r(s)\equiv \frac{1}{r} \Psi(\frac{s}{r})$.  If the scale $r$ is
small enough, the integral is dominated by the first terms in the
pseudo-Taylor expansion of $F(\vec{x})$ around $\vec{x}_0$,
eq.~(\ref{eq:pseudoTaylor}):

\eq
T^{\vec{r}}_{\Psi} F(\vec{x}_0)\;\approx\;  \int ds \; p_n(s)
\Psi_r(s)\; + \; A_0 \int ds \; |s|^{h_F(\vec{x}_0)} \Psi_r(s)
\en

\noindent
If the analyzing wavelet is now required to vanish at least the 
first $n$ integer moments,  
the singular part appears as the main contribution to the wavelet 
transform. (For later reference, let us say now that
a wavelet with a vanishing zero-th order moment -- i.e., its 
integral is null -- is called an \emph{admisible wavelet}).  Hence, 
changing variables $t=\frac{s}{r}$ and defining $A=A_0\: \int
dt\; |t|^{h_F(\vec{x}_0)} \Psi (t)$, one has:

\eq
T^{\vec{r}}_{\Psi} F(\vec{x}_0)\;\approx\; A \: r^{h_F(\vec{x}_0)} 
\label{eq:WT_scaling}
\en

\noindent
where all the dependence on $r$ stands in the prefactor 
$r^{h_F(\vec{x}_0)}$. In this way, the H{\"o}lder exponent (in the sense of
eq.~(\ref{eq:Holder_general})) can be obtained by a wavelet
projection, provided the wavelet has zero moments up to a large enough
order. Since $h_F(\vec{x}_0)$ is non-integer, this procedure cannot 
cancel the singular part even if the order $n$ of the largest vanishing 
moment of $\Psi$ is larger than $h_F(\vec{x}_0)$. On the other hand, the 
wavelet could not be able to detect singularity exponents $h_F > n+1$.  
This is because a polynomial of order $n+1$ could still be present and 
dominate the contribution of the singular term. For this reason, it is 
important to determine the right order of the required zero moments of 
the wavelet.

\indent
This is then the appropriate scheme to detect the H{\"o}lder
exponents of signals with a regular part. 
In cases in which the $\mu $-density $|\nabla C|$
and/or its primitive $C$ are not multiaffine,  
the concept of multiaffinity (eq.~(\ref{eq:Holder})) 
should be applied to their wavelet projections. 
It is then necessary to determine how many zero moments  
the wavelet should have, as it provides information on the global aspects of
the signal. This generalization is straightforward, replacing in all the
cases $\delta_{\vec{r}} C$ by $T^{\vec{r}}_{\Psi} C$ 
and $\delta_{\vec{r}} |\nabla C|$ by $T^{\vec{r}}_{\Psi} |\nabla C|$. 
For instance, if the wavelet transforms of $C$ and $|\nabla C|$ verify
eq.~(\ref{eq:WT_scaling}) for certain $A$ and $h_F(\vec{x})$, the same 
arguments that led us to eqs.~(\ref{eq:relation_LIs}) and 
(\ref{eq:relation_LIs_taup}) can be used to derive that~:

\eq
\epsilon_r\; \sim  \; T^{\vec{r}}_{\Psi} |\nabla C|\; \sim\; \frac{1}{r}\:  
T^{\vec{r}}_{\Psi}  C 
\label{eq:relation_WT_LI}
\en

\noindent
which is expressed in terms of the SS exponents $\tau_p^{\Psi_{_C}}$ (of 
$T^{\vec{r}}_{\Psi}  C$) and $\tau_p^{\Psi_{_{\nabla C}}}$ (of 
$T^{\vec{r}}_{\Psi} |\nabla C|$) as:

\eq
\tau_p^{\Psi_{_C}}\; =\;  \tau_p^{\Psi_{_{\nabla C}}} + p \; = \; \tau_p +p
\label{eq:relation_WTs_taup}
\en

\indent
Let us make a final remark: it is possible that the wavelet
projections of a given scalar function $F$ verifies the scaling in
eq.~(\ref{eq:WT_scaling}) without verifying
eq.~(\ref{eq:Holder_general}). This can happen, for instance, for
functions with such an irregualar behavior that the pseudo-Taylor
expansion is not possible. In those cases, $T^{\vec{r}}_{\Psi} F$ can
be used as a stronger generalization of the usual Linear Increment.  As 
it will be shown in Section \ref{subsection:exp.multiaffine}), this is 
the case for $|\nabla C|$.


\section{Numerical analysis of images}
\label{section:results}

\subsection{Databases}
\label{subsection:exp.database}

\indent

We have used two different image datasets: one from Daniel Ruderman
(see \cite{Rud94}), consisting in 45 B/W images of $256\times 256$ pixels
with up to 13 bits in luminosity depth (from now on this set will be
referred to as the first dataset); and another containing 25 images
taken from Hans van Hateren's database (see \cite{vanH+98}),
having $1536\times 1024$ pixels and 16 bits in luminosity depth (we will
refer to these images as the second dataset).  These 25 images are
enumerated in table \ref{table:vH_images}

\begin{table}[htb]
\begin{tabular}{|c||c||c||c||c|} \hline
imk00034.imc & imk00801.imc & imk01164.imc & imk02649.imc & imk03807.imc \\ 
imk00211.imc & imk00808.imc & imk01406.imc & imk03134.imc & imk03842.imc \\ 
imk00478.imc & imk00881.imc & imk02035.imc & imk03514.imc & imk03940.imc \\ 
imk00586.imc & imk01017.imc & imk02603.imc & imk03536.imc & imk04042.imc \\ 
imk00605.imc & imk01032.imc & imk02626.imc & imk03789.imc & imk04069.imc \\ \hline
\end{tabular}
\caption{
The 25 images from van Hateren's database used  
in this study. They can be observed and downloaded from the URL address    
http://hlab.phys.rug.nl/archive.html
} 
\label{table:vH_images} 
\end{table}

\indent 
The first dataset has been used for the statistical analysis of the
one-dimensional wavelet transform, eq.~(\ref{eq:1d_restriction}), while
the second dataset was used for the more demanding task (from the
statistics point of view) of computing the moments of the
two-dimensional variable in eq.~(\ref{eq:square}). No matter the
ensemble considered, however, the studied properties seem to behave in
essentially the same way.

\subsection{Multifractality of the measure and singularity analysis}
\label{subsection:exp.wavelet}

\indent
We first address the issue of the multifractality of the measure $\mu $
defined in eq.~(\ref{eq:measure}). Our numerical analysis gives an
affirmative answer to this question: the wavelet transform
$T^r_{\Psi}d\mu (\vec{x}_0)$ is well fitted by $\sim r^{h(\vec{x}_0)}$ in
about $98\%$ of the pixels $\vec{x}_0$ in each image of our two
datasets.  The analysis was done using several wavelets, although the
most convenient family of one parameter functions was
$\Psi^{\gamma}(\vec{x})=\frac{1}{(1+|\vec{x}|^2)^\gamma}$ ($\gamma
\geq 1$). These are positive (non-admissible) wavelets.  The
singularity exponent of the measure at a given pixel, $h(\vec{x})$,
was obtained from a regression on eq.~(\ref{eq:exponent_wv}).  The
value of the scale $r$ was typically taken in the range $(0.5, 2.0)$
in pixel units. Notice that for scales smaller than $r \sim 0.5$ the
method would detect an artificial ``edge'' that actually corresponds
to the pixel boundaries, what corresponds to $h=-1$. Similarly, as the
scale becomes large the wavelet can only discriminate between very
sharp edges and, for $r$ of the order of the size of the image it
would yield the value $h=0$ everywhere.

\indent
To obtain the fractal components $F_h$, the pixels of a given image
were classified according to the value of their exponents.  The
exponents lie in the interval $(-0.5,1.4)$: only a negligible fraction
of pixels have exponents less than $-0.5$ (~about $0.8 \%$ of them~)
and none with a singularity below $-1$. It is also observed that there
are no exponents larger than $1.4$. This range of possible exponent
values coincides exactly with that predicted in \cite{Tur+98}, and
is in agreemente with the model explained in section
\ref{subsection:ssandmulti}.

\begin{figure}[htb]
\vspace*{-1cm}
\begin{center}
\leavevmode
\epsfxsize=8cm
\epsffile{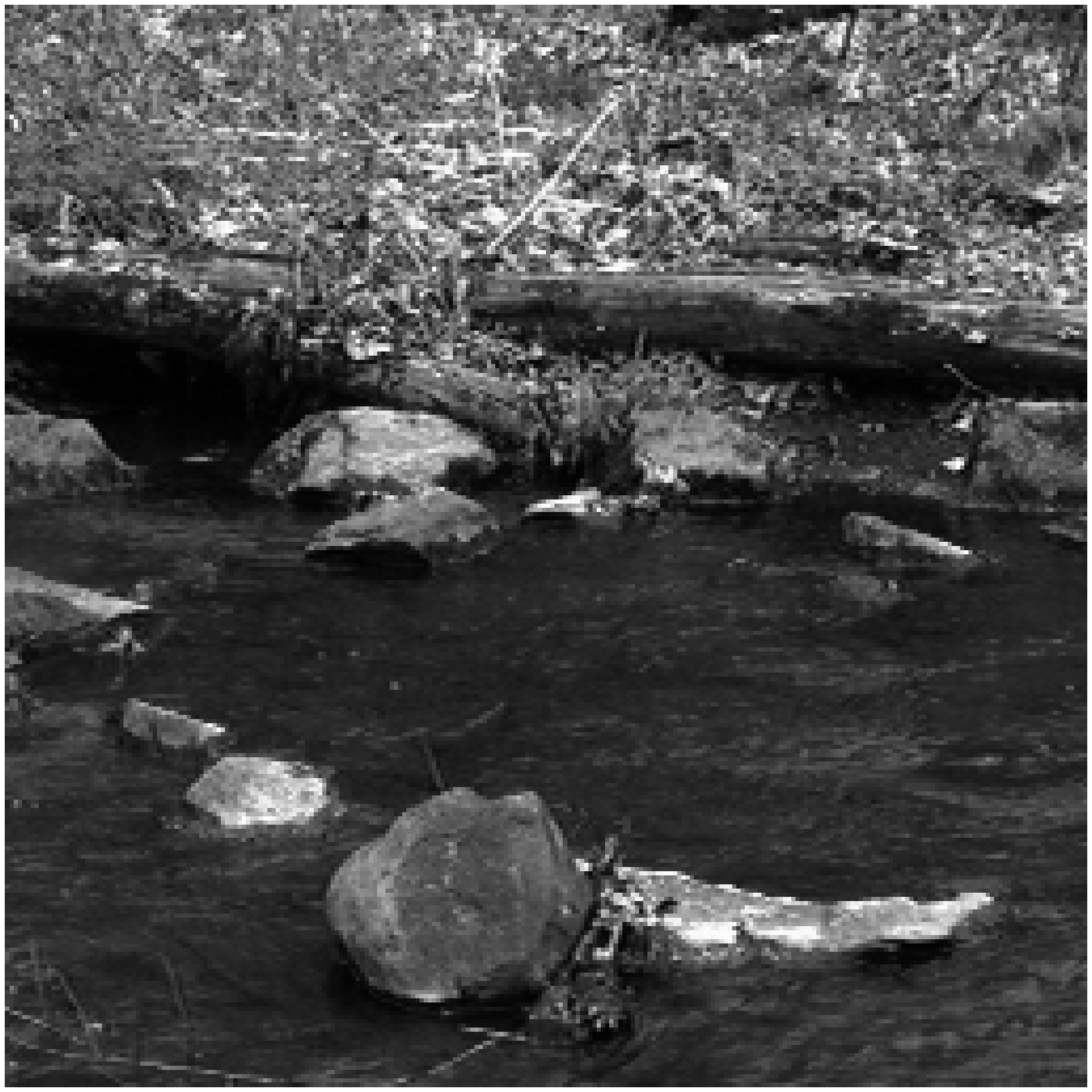}
\\
\vspace*{0.3cm}
\leavevmode
\epsfxsize=8cm
\epsffile{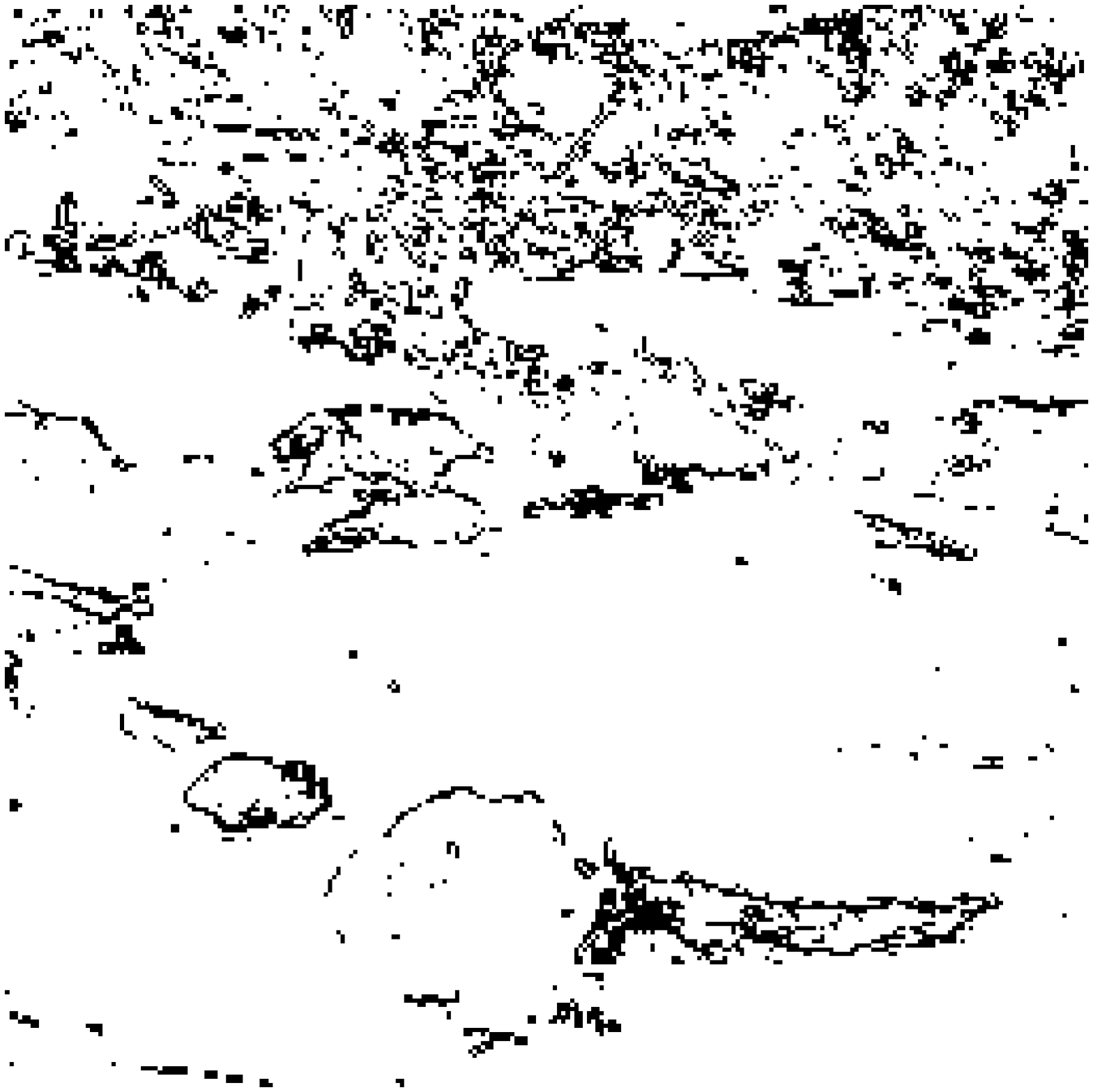}
\\
\end{center}
\caption{
An image from Ruderman's ensemble
\protect\cite{Rud94} and its most singular component (taken as the set of
points having a singularity exponent in the interval $h=-0.5\pm 
0.05$).
}
\label{fig:most_singular}
\end{figure}

\indent 
The most singular component is the one with $h=-0.5$, defined within a
window of size $\Delta h$. The visualization of this set is very
instructive (see Fig.~\ref{fig:most_singular}).  Clearly, the points
appear to be associated with the contour of the objects present in the
image. This is in agreement with our expectation that the behavior of
the measure at these edges should be singular, and as much as
possible. But there is something rather surprising: for a sharp,
theta-like contrast, with independent values at both sides of the
discontinuity, one would have expected an exponent $-1$, which appears
as a Dirac $\delta$-function in the density measure $\nabla C$ (see
example number 4 in the Appendix).  Our interpretation of the minimal
value of $h$, $h_{\infty}=-0.5$, is that it reflects the existence of
correlations among different fractal components of the images that
smoothen the singularities (in particular those of the sharpest
edges).

\indent 

\begin{figure}[htb]
\vspace*{-1cm}
\begin{center}
\leavevmode
\epsfxsize=8cm
\epsffile{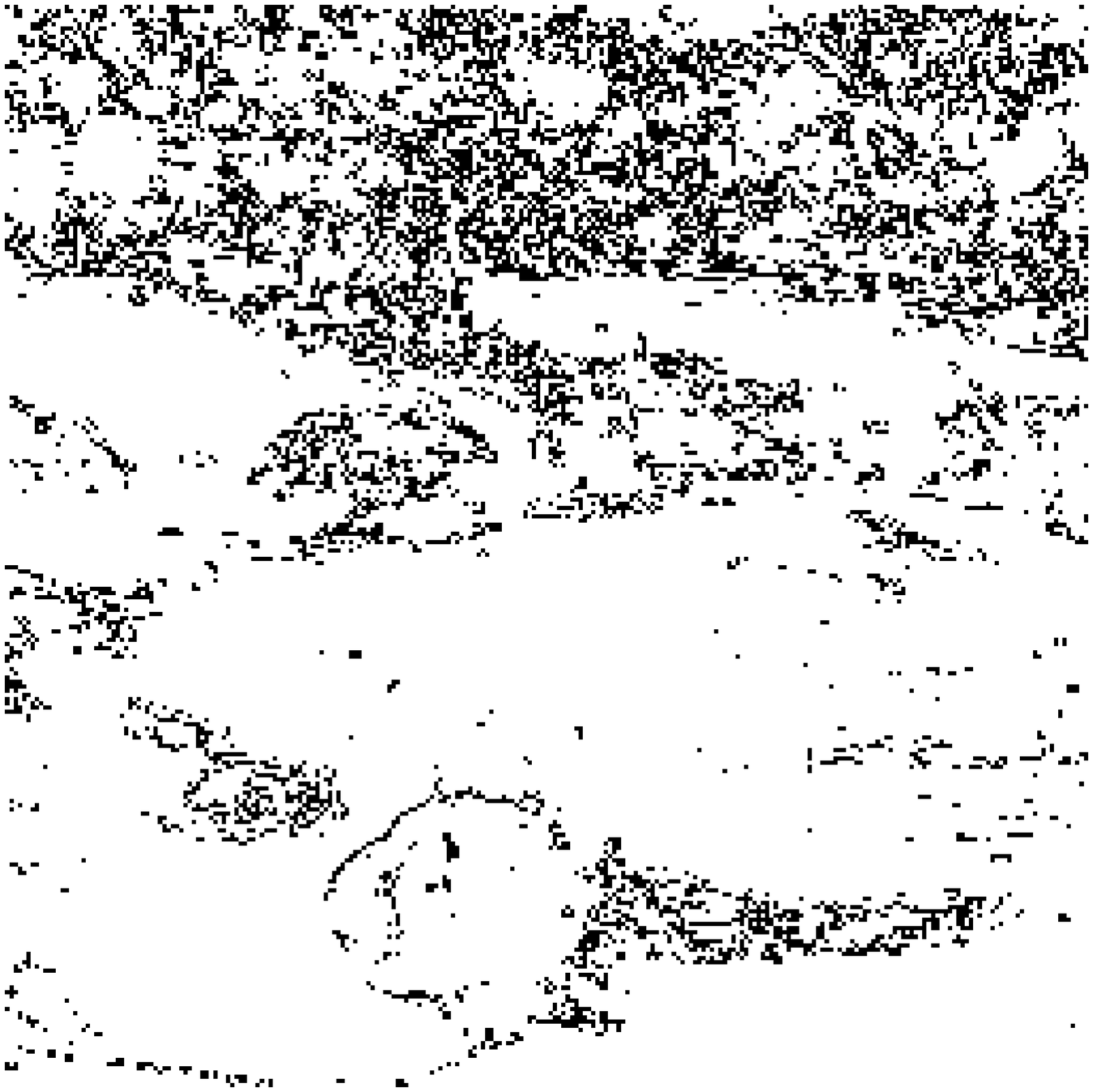}
\\
\vspace*{0.3cm}
\leavevmode
\epsfxsize=8cm
\epsffile{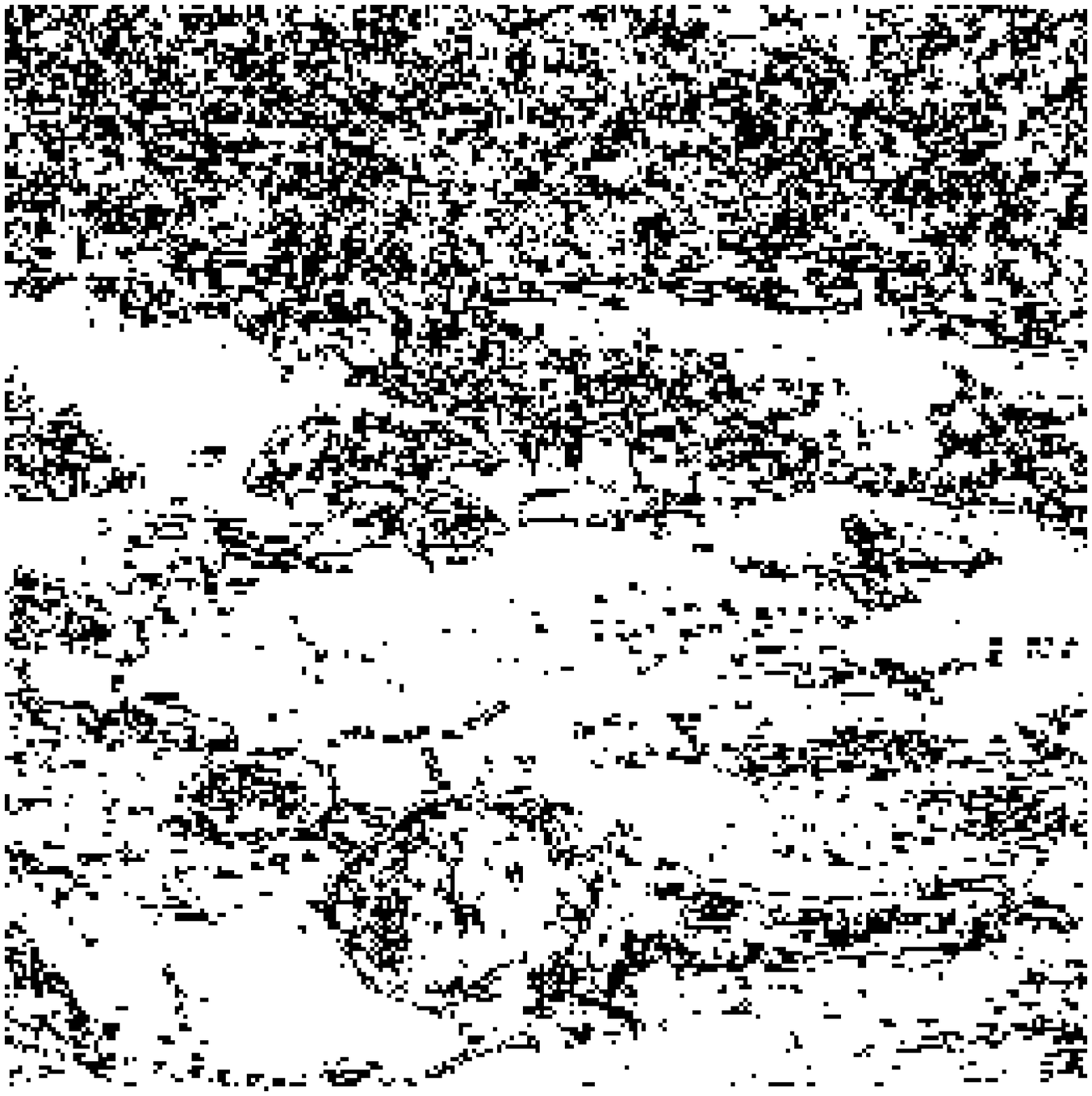}
\end{center}
\caption{
Fractal components with exponents $h=-0.4\pm 0.05$ (above) and
$h=-0.3\pm 0.05$ (below) for the same image represented in Fig.
\ref{fig:most_singular}. Notice the spatial correlations between
components with similar singularities. 
}
\label{fig:other_manifolds}
\end{figure}

It is also interesting to observe the fractal components with
exponents just above the most singular
one. Fig. \ref{fig:other_manifolds} shows the two next components,
defined with $\Delta h = 0.05$. A comparison with Fig.
\ref{fig:most_singular} shows that pixels in these components are
close to pixels in the most singular one.  It is also rather clear
that the probability that a random ball of size $r$ contains a pixel
from the component $F_h$ increases with $h$, at least for small values
of this exponent. Since this probability scales as shown in
eq.~(\ref{eq:component_prob}), this behavior reveals an increase with
$h$ of the dimensionality of the fractal components. It is also
observed (although not shown in the figures) that the dimensionality
starts decreasing beyond $h\approx 0.2$. This is in agreement with the
arguments of section \ref{subsection:ssandmulti}: the singularity
spectrum $D(h)$ of the S-L model, which is shown in
Fig.~\ref{fig:Dh_S-L}, reflects this behavior of the fractal
components.

The analysis done in this subsection justifies the arguments presented
in Section \ref{section:introduction}.  One cannot
consider the sharpest edges of an image independently of the other
texture structures (i.e. the other fractal components). Luminosity jumps
from one roughly constant value to another (statistically independent)
constant value are not typical in natural images. The sharpest edges
appear correlated with smoother textures, an effect that tends to
decrease the strentgh of their singularity exponent, reaching the
value $h \sim -0.5$.  Standard edge-detection tecniques  
\cite{Gon+92} have not been
designed to search for singularities of this strength. Good
edge-detection algorithms should be specifically constructed to
capture singularities close to $h_{\infty}$.

\subsection{Statistical analysis of the measure}
\label{subsection:exp.SS}

\indent
We computed the $p$-moments of three different variables 
related to the measure:

\begin{itemize}

\item
{\bf Bidimensional Edge Content ($\epsilon_r$):} Following the definition given
in eq.~(\ref{eq:square}). This analysis was done using the
25 images of the second dataset. As the variable itself is positive,
we computed directly the quantities $\langle \epsilon_r^p \rangle$.

\item
{\bf Two one-dimensional restrictions of the wavelet projection of 
the measure\footnote{ 
Since the calculations involving $T^r_{\Psi} d\mu $ are
highly computer-time consuming we did not consider the bi-dimensional
wavelet transform.  FFT cannot be safely used due to aliasing,
which heavily changes the tails (rare events) of the distribution.  }
($T^r_{\Psi} d\mu $):} We considered horizontal ($h$) and vertical
($v$) restrictions of the measure $d\mu $, that is $d\mu ^h=dx
|\frac{\partial\mu }{\partial x}|$ and $d\mu ^v=dy
|\frac{\partial\mu }{\partial y}|$. The appropriate wavelet transforms
are denoted by $T^r_{\Psi} d\mu ^l$ ($l=h,v$). As these coefficients
need not to be positive (in general $\Psi$ is not positive), we
computed the moments of their absolute values, $\langle
|T^r_{\Psi}d\mu ^l|^p\rangle$.
 
\end{itemize}

\begin{figure}[htb]
\begin{center}
\hbox{
	\hspace*{-1cm}
	\makebox[0.5cm]{$\rho(p,2)$}
	\makebox[12cm]{\hspace*{0.25cm}$a$\hspace*{3.75cm}$b$\hspace*{3.75cm}$c$}
}
\vspace*{.1cm}
\hbox{
	\hspace*{-1cm}
	\makebox[0.5cm]{}
	\makebox[12cm]{
		\leavevmode
		\epsfxsize=4cm
		\epsfysize=4.5cm
		\epsfbox[115 60 363 292]{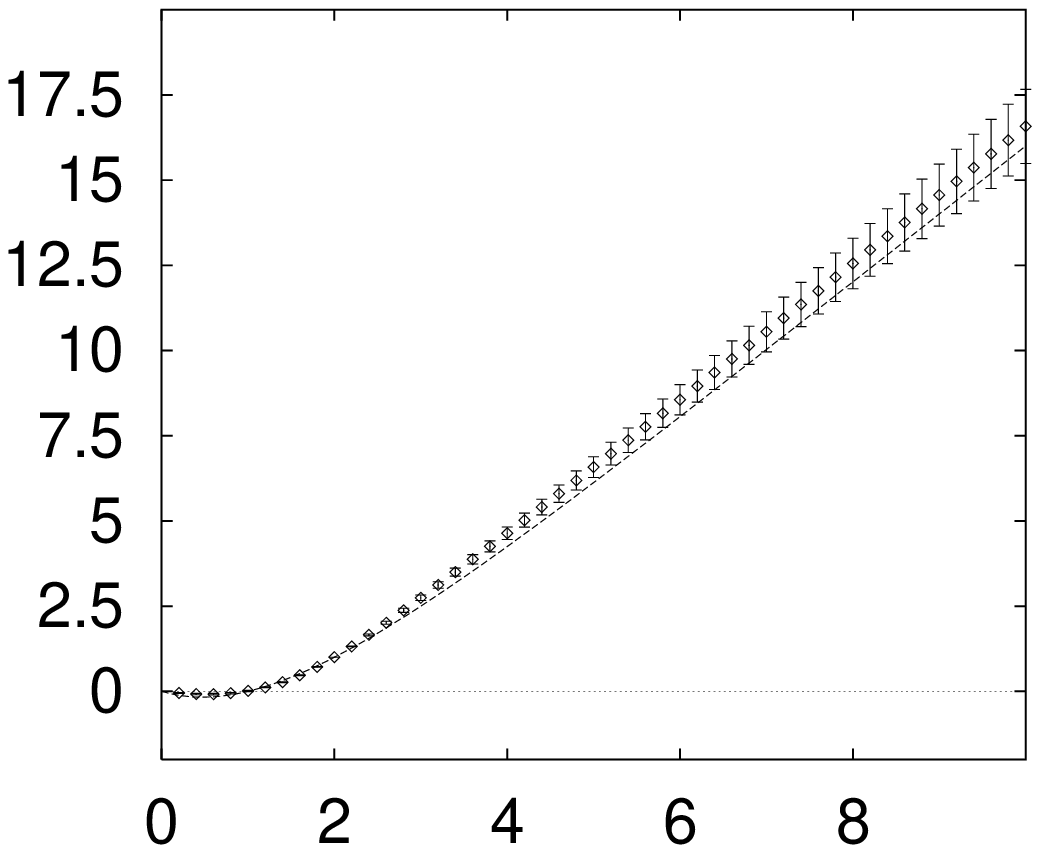}
		\epsfxsize=4cm
		\epsfysize=4.5cm
		\epsfbox[115 60 363 292]{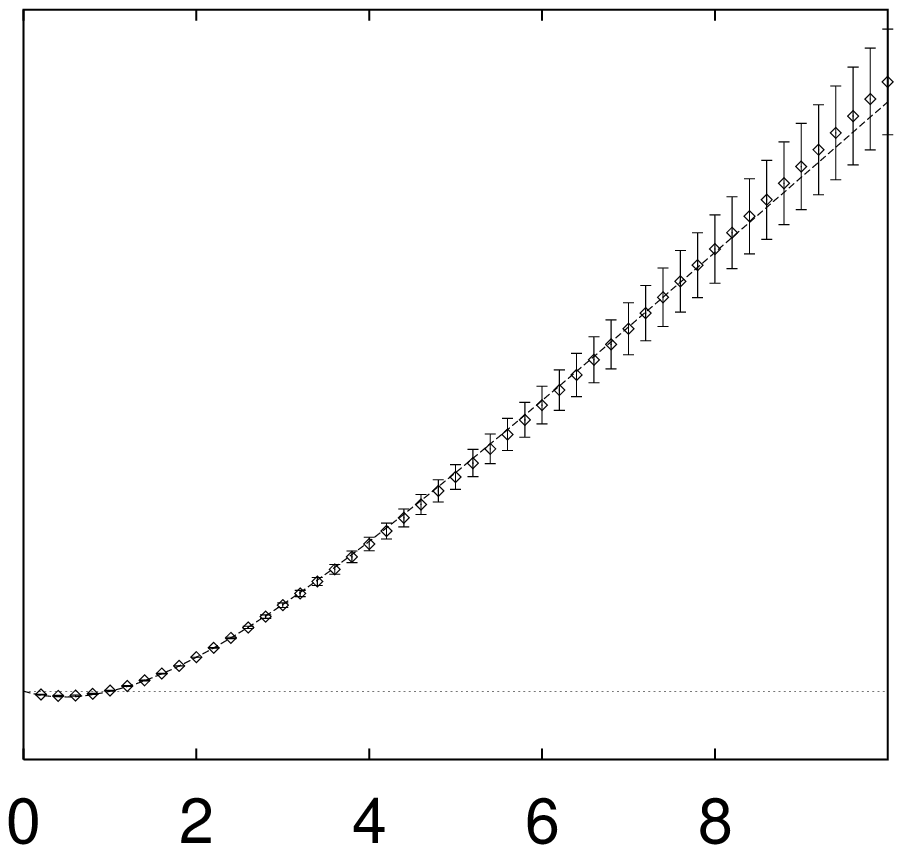}
		\epsfxsize=4cm
		\epsfysize=4.5cm
		\epsfbox[115 60 363 292]{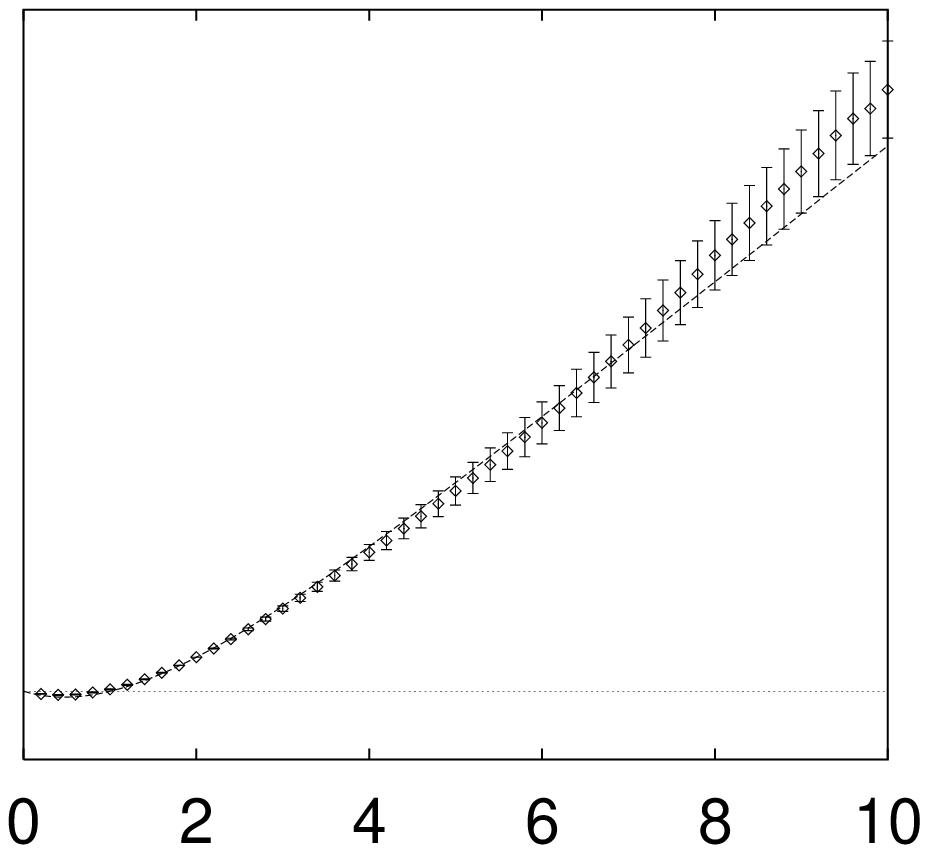}
	}
}
\hbox{ 
	\hspace*{-1cm}
	\makebox[0.5cm]{}
	\makebox[12cm]{\hspace*{0.25cm}$p$\hspace*{3.75cm}$p$\hspace*{3.75cm}$p$} 
}
\vspace*{.5cm} 
\caption{ 
ESS exponents $\protect\rho(p,2)$ for {\bf a)} the
bidimensional edge content $\epsilon_r$ , {\bf b)} the horizontal wavelet
transform $T^r_{\Psi}d\mu ^h$  and {\bf c)} the vertical one,
$T^r_{\Psi}d\mu ^h$. The convolution function  
$\Psi$ was taken to be a gaussian function. Each value of
$\protect\rho(p,2)$ was obtained by linear regression of the logarithm of
the $p$-th moment versus the logarithm of the second moment, for $r$
between 8 and 32. The solid line represents the fit with the Log-Poisson
process. The best fit is obtained with {\bf a)} $\protect\beta = 0.52\pm 0.05$
{\bf b)} $\protect\beta = 0.55\pm 0.06$ and {\bf c)} $\protect\beta = 0.5\pm 0.07$
The SS parameters $\tau_2$ were also calculated; they turned out to be 
{\bf a)}  $\tau_2=-0.26\pm 0.06$ , {\bf b)}  $\tau_2=-0.25\pm 0.06$ and
{\bf c)} $\tau_2=-0.25\pm 0.08$. 
}  
\label{fig:rho-tau} 
\end{center}
\end{figure}

\indent
It is remarkable that the bi-dimensional Edge Content and the two
one-dimensional wavelet projections exhibit the scaling properties of
SS (eq.~\ref{eq:SS}) and of ESS (eq.~\ref{eq:ESS}) (the ESS tests for
the three variables are presented in Fig.~\ref{fig:ESS-epsilon} and
\ref{fig:ESS-grad_wv}).  This was also observed in \cite{Tur+98},
although for somewhat different variables: the one-dimensional
horizontal and vertical restrictions of the local edge variance.
Surprisingly, no matter the particular
variable, the exponents $\tau_p$ and $\rho(p,2)$ obtained are very
similar (see Fig.~\ref{fig:rho-tau} and those in \cite{Tur+98}).
Therefore, they should refer to very essential, robust aspects of 
images.

\indent
The S-L model fits very well the experimental exponents (see  
Fig.~\ref{fig:rho-tau}). It seems then that the multifractal structure 
underlying the statistical description could be described with the 
Log-Poisson process, which means that under infinitesimal changes in the 
scale there are only two possible different types of transformations.

\indent 
It is really remarkable that for our datasets $\Delta\approx 0.5$ 
and $D_{\infty}\approx 1$. This means that the most singular
component is a collection of lines (the sharp edges present in the
image) which can be characterized by their common exponent
$-\Delta=-0.5$. But this is precisely the smallest value we observed
previously in our wavelet analysis. The statistical approach confirms
the result obtained by a local singularity analysis.

The non-linear dependence of $\rho(p,2)$ on $p$ again indicates that
the Edge Measure is multifractal. This has to be contrasted with the
results by \cite{Rud+94,Rud94} who find that the log-contrast
distribution follows a simple classical scaling.
\footnote{
Other luminosity analysis performed by D. Ruderman provided
some evidence of multiscaling behaviour (private communication).}
It also holds for any derived variable as the gradient (see \cite{Rud94}
for
details). This particular scaling is related, in that context, with
general properties of scale invariance of natural images: the averaging of
the contrast does not keep track of the details (e.g., the edges
completely contained into the averaging block) and produces a
statistically equivalent image. The Edge Content and the wavelet
projections, being local averages of a gradient, are able to keep the
information about the more complex underlying multifractal structure. 

The same authors have also designed an iterative procedure that tends
to decompose the image into a gaussian-like piece (local fluctuations
of the log-contrast) and an image-like piece (local variances). Such a
decomposition would suggest the existence of a multiplicative process
for the log-contrast itself where the random variable relating the
image at two different levels of the hierarchy is given by the local
fluctuations.  A closer look to the iterative procedure indicates that
this is not the case: independence between the two factors of the
process is not guaranteed and also the process is not infinitely
divisible.

\subsection{Multiaffinity properties of $C$ and $|\nabla C|$}
\label{subsection:exp.multiaffine}

\indent

In this section we study the scaling properties for the Linear Increments of
the contrast $C(\vec{x})$ and of the measure density $|\nabla
C|(\vec{x})$. Our aim is then to check if $C$ and $|\nabla C|$ 
satisfy the scaling defined in eq.~(\ref{eq:multiaffine}). 
Recalling the connection (the analog of eq.~(\ref{eq:preSP}) for 
these two quantities) between 
local scaling (described by the singularity exponents) and  
statistical scaling (characterized by the SS exponents), 
one concludes that the validity of SS implies that 
multiaffinity holds. We have then done a statistical 
analysis of the Linear Increments of the contrast and the measure density.

\begin{figure}[htb]
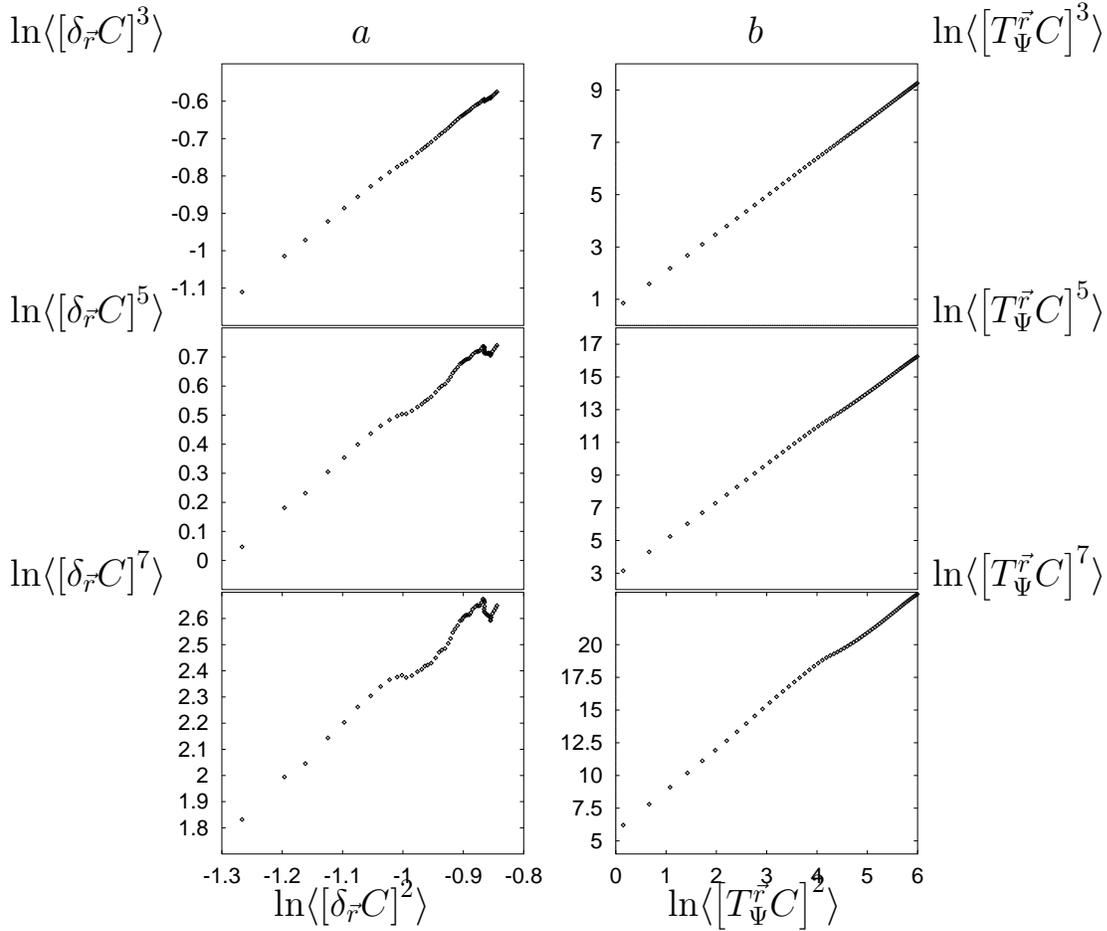

\caption{
Test of ESS for {\bf a)} the horizontal linear increment of the contrast, 
$\delta_{r\vec{i}}C$ and {\bf b)} the horizontal wavelet coefficient 
$|T^r_{\Psi} C^h|$ ($\Psi$ was taken as the second derivative of a 
gaussian); the first dataset was used and the range of scales was 
from $r=4$ to $r=64$. ESS is not present for $\delta_{r\vec{i}}C$, but 
it is clearly observed for the wavelet transform  (notice the presence of 
the cut-offs due to numerical effects). 
Note that the scale of the axes in {\bf a)} and {\bf b)} is different.
} 
\label{fig:ESS-contLI}
\end{figure}

\indent 
We first computed the $p$-moments of the Linear Increments of both
variables, looking for SS and ESS scalings. Let us notice that there
is no {\it a priori} argument to expect that $\delta_{\vec{r}}
C(\vec{x})$ shows SS. On the other hand, since the measure $\mu $ is
multifractal (as it was shown in section \ref{subsection:exp.wavelet})
it would be reasonable to expect that $|\nabla C|(\vec{x})$ has a
multiaffine behavior. However, this has to be verified explicitly
because the multifractality of the measure does not guarantee the
multiaffinity of its density, as it was discussed in section
\ref{subsection:multiaffine}.  In fact, our numerical analysis showed
that neither $C(\vec{x})$ nor $|\nabla C|(\vec{x})$ have SS. The same
is true for ESS, as it is shown in Figs.~\ref{fig:ESS-contLI}a and
~\ref{fig:ESS-gradLI}, where the moments exhibit a rather erratic behavior.


\indent
It seems that the failure of SS for $C(\vec{x})$ and $|\nabla C|(\vec{x})$ 
has to be explained in different terms.
In the case of $|\nabla C|(\vec{x})$,  
this quantity is the measure density and, 
as it was stated in eq.~(\ref{eq:exponent_wv}),
its convolution with a non-admissible (i.e. positive) 
wavelet has the same singularities as the measure, 
eq.~(\ref{eq:measure}). This implies that the lack of SS 
cannot be explained 
in terms of a polynomial part as described in 
eq.~(\ref{eq:pseudoTaylor}). In fact, a 
positive wavelet cannot alter the contribution from this 
polynomial while one observes that the data for  
$\delta_{\vec{r}} |\nabla C|(\vec{x})$ (Fig.~\ref{fig:ESS-gradLI}) 
are different from those of 
$T^r_{\Psi} d\mu ^h$ and $T^r_{\Psi} d\mu ^v$
(Fig.~\ref{fig:ESS-grad_wv}).

On the other hand it is plausible that the 
failure of SS of $\delta_{\vec{r}} C(\vec{x})$ can be 
attributed to the presence of a polynomial part in 
$C(\vec{x})$. Differently to what happened with $|\nabla C|(\vec{x})$, 
when the contrast is convolved with a non-admissible 
wavelet, SS and ESS are still not present (that is, one 
observes a similar erratic behavior as in Fig.~\ref{fig:ESS-contLI}a). 
The same behavior is observed when $C(\vec{x})$ 
is convolved with an admissible wavelet with a zero mean. 
The situation changes drastically when a wavelet with 
vanishing zero- and first-order moments is used. 
The result obtained using the second derivative 
of a gaussian is presented in Fig.~\ref{fig:ESS-contLI}b, 
which clearly shows ESS (it also exhibits SS).   
This can be understood recalling that the possible 
values of the H{\"o}lder exponent of $C(\vec{x})$
are contained in the interval $(0.5, 2.4)$ 
(see eq.~(\ref{eq:shift-1}) and section \ref{subsection:exp.wavelet}).
A non-admissible wavelet truncates the detected singularities at the 
value $h=0$, an admissible one truncates them at $h=1$. 
The third wavelet considered, with two vanishing moments, 
truncates the singularities only from $h=2$ and then it covers almost the 
whole interval (In theory a wavelet with another vanishing moment  
would be more appropriate, however in practice it 
is numerically more unstable because the resolution of the 
wavelet is smaller).

\indent
Finally, we also checked that the experimental values of $\tau_p$, 
$\tau_p^{\Psi_{_C}}$ and $\tau_p^{\Psi_{_{\nabla C}}}$ satisfy the relation 
given in eq.~(\ref{eq:relation_WTs_taup}).


\section{Discussion}
\label{section:discussion}

We have shown that in natural scenes pixels are arranged in
non-overlapping classes, the fractal components of the image,
characterized by the behavior of the contrast gradient under changes
in the scale.  In each of these classes this quantity exhibits a
distinct power law behavior under scale transformations, what
indicates that there is no a well-defined scale for the isolated
components but that at the same time the scene is not scale invariant
globally.

\indent 
This result implies that a given image is composed of many
different spatial structures of the contrast associated to the fractal
components, which can be arranged in a hierarchical way. In fact, how
sharp or soft a change in contrast is at a given point can be
quantified in terms of the value of the scaling exponent at that
site. The smaller the exponent the sharper the local change in
contrast is.

\indent
More precisely, we have dealt with two basic quantities: the contrast
$C(\vec{x})$ and the edge measure $\mu $, which density is the modulus of
the contrast gradient $|\nabla C|(\vec{x})$ (see
eq.~(\ref{eq:measure})).  Closely related to the measure, we can
define the edge content of a region $A$ of the image
(eq.~(\ref{eq:edge_content})), which quantifies how much the contrast
gradient deviates from being uniformly distributed inside the region
$A$. This is a suitable definition to study the local behavior of the
contrast gradient.  As the size of $A$ is reduced, some contributions
of the contrast gradient to the edge content are left outside this
region. Under an infinitesimal change in the scale very often these
contributions are small, but sometimes it does happen that a small
change in the scale produces a large change in the edge content. This
is due to singularities in the contrast gradient. As a consequence of
this, the edge content (and the measure) exhibits large spatial
fluctuations. The edge content can have different singularities at
neighbouring points, characterized by a power-law behavior with
exponents $h(\vec{x})$ which depend on the site,
eq.~(\ref{eq:exponent_epsilon}). This means that the measure is
multifractal, as it has been explicitly verified in this work.

\indent 
The set of pixels with the same value of $h(\vec{x})$ defines a
particular spatial structure of contrast: the fractal component
$F_h$. The smallest $h(\vec{x})$ is the sharpest the contrast gradient
at the pixel $\vec{x}$ is.  Its minimal value was, for our dataset,
$h_{\infty} = -0.5$. Another question refers to the fractal dimension
$D_{\infty}$ of this component. One finds $D_{\infty} = 1$
\cite{Tur+98}, and when this component is extracted from the image
one checks that it corresponds to the boundaries of the objects in the
scene. The associated contrast structure is made of the sharpest edges
of the image. The other, less singular components represent softer
contrast structures inside the objects. It is also observed that there
are strong spatial correlations between fractal components with
similar exponents.

\indent
We have also shown how to extract the fractal components from the
image. To do this one needs a suitable technique to compute the scale
behavior at a given pixel.  A wavelet projection
\cite{Mall+92,Arn96} seems to be the correct way to perform this
analysis, since it provides a way of interpolating scale values which
are not an integer number of pixels.  Our approach differs from the
one followed in those references in several aspects, but in particular
in that we have been here interested in the characterization of the
whole set of fractal components.

\indent
We remark the important fact that the multifractal can be explained by
a model with only two parameters (e.g. $\beta$ and $\Delta$). This had
been noticed before in terms of a variable that integrates the square
of the derivative of the contrast along a segment of size $r$, both
taken either in the horizontal or in the vertical direction
\cite{Tur+98} .  This quantity admits the interpretation of a
local linear edge variance.  We have seen that the same model is valid
for many other quantities, some of them more general and natural.

\indent 
The basic idea of the model is that the contrast gradient has
singularities that under an infinitesimal change in the size of $A$
produce a substantial change in the edge content inside that region.
The simplest stochastic process to assign a value to this change is a
binomial distribution, eq~(\ref{eq:Log-Poisson_inf}): its more likely
value is very small, but with small probability this change is
given by the parameter $\beta$, which for our dataset is about
$0.5$. The other parameter of the model, $\Delta$, measures the
strength of the singularity. The geometrical locus of these
singularities is the most singular fractal component. For a finite scale
transformation this process becomes Log-Poisson. The model appeared
before in other problems \cite{She94}.

\indent
In a sense, the most singular component conveys a lot of information
about the whole image. The two model parameters can be obtained by
analyzing properties of this particular class of pixels: its dimension
($D_{\infty}$) and the scale exponent of the edge content
($\Delta=-h_{\infty}$).  In turn, these two parameters completely
define the whole dimension spectrum in the Log-Poisson model. It is
then very plausible that other components have a great deal of
redundancy with respect to the most singular one.

\indent
This hierarchical representation of spatial structure can be used to
predict specific feature detectors in the visual pathways. We
conjecture that the most singular component contains highly
informative pixels of the images that are responsible for the
epigenetic development leading to the adaptation of receptive fields.
Learning of the statistical regularities \cite{Bar61} present in
this portion of visual scenes (using for instance the algorithm in 
\cite{Bel+97}), would give relevant predictions
about the structure of receptive fields of cells in the early visual
system.



\section*{Acknowledgments}

We are thankful to Jean-Pierre Nadal, Germ{\'a}n Mato and Dan Ruderman for
useful comments and to Angel Nevado for many fruitful discussions that
we had during the preparation of this work.  We are grateful to Dan
Ruderman and Hans van Hateren, some of whose images we used in the
present statistical analysis.  Antonio Turiel is financially supported
by a FPI grant from the Comunidad Aut{\'o}noma de Madrid, Spain.  This
work was funded by a Spanish grant PB96-0047.

\newpage

\appendix

\section{Characterization of the singularities of simple functions}

\indent
It is instructive to compute the singularity exponents defined in
eq.~(\ref{eq:Holder}) for several simple functions.

\begin{enumerate}

\item
{\bf Continuous functions:} If a function $f(\vec{x})$ is continuous at a 
given point $\vec{x}$, the value of $f$ at $\vec{y}$ should be 
rather close to $f(\vec{x})$ provided $\vec{y}$ is close to $\vec{x}$; 
then $|f(\vec{y})|<|f(\vec{x})|+\epsilon$ so  
$|f(\vec{x})-f(\vec{y})|<2|f(\vec{x})|+\epsilon=A$. That is, 
\underline{$f$ is H{\"o}lder of exponent $0$ at $\vec{x}$.}

\item
{\bf Smooth functions with continuous, non-vanishing first derivative:} Using 
the Taylor expansion, $f(\vec{x})-f(\vec{y})= \nabla f(\vec{y}_0) \cdot 
(\vec{x}-\vec{y})$ with $\vec{y}_0$ a point between $\vec{x}$ and $\vec{y}$.
Let us call $A=\max |\nabla f(\vec{y}_0)|>0$, so 
$|f(\vec{x})-f(\vec{y})|< A |\vec{x}-\vec{y}|$. That is, \underline{$f$ is 
H{\"o}lder of exponent $1$ at $\vec{x}$.}

\indent
Analogously, any function $f$ having $n-1$ vanishing derivatives at a 
point $\vec{x}$ and a non-vanishing continuous $n$-th derivative,
is H{\"o}lder of exponent $n$ at $\vec{x}$.

\item
{\bf $\theta$-function:} The $\theta(x)$ function is given by 

\[
\theta(x) \; =\; \left\{ \begin{array}{cc}
0 & x<0\\
1 & x>0\\
\mbox{undefined} & x=0
\end{array}\right.
\]

\indent
Since for $x\neq0$ the increments are zero (for  
small enough displacement) it is H{\"o}lder of any order. 
For $x=0$ we make use of the 
property that $f$ is H{\"o}lder of exponent $h$ at a given point if and 
only if its primitive $F$ is H{\"o}lder of exponent $h+1$ at the same point.

\indent
One possible primitive $F(x)$ of $\theta(x)$ is the following: 

\[
F(x) \; =\; \left\{ \begin{array}{cc}
0 & x<0\\
x & x\geq 0\\
\end{array}\right.
\]

\noindent
which is a continuous function. The increments $|F(0)-F(r)|=|F(r)|=F(r)$ are 
thus immediate, and obviously $F(r)<|r|$. Moreover, it is also 
clear that the maximal $h$ such that $|F(0)-F(r)|<A |r|^h$ is $h=1$. 
Thus, $F$ has H{\"o}lder exponent 1 at $x=0$ and therefore 
\underline{$\theta(x)$ has H{\"o}lder exponent 0 at $x=0$.}

\item
{\bf $\delta$-function:} The Dirac's $\delta$-function is in fact a 
distribution. Anyway, it is possible to define the H{\"o}lder exponent of  
distributions attending to the fact that any distribution can be expressed 
as a finite order derivative of a bounded function. So, if the $n$-th 
order primitive of a distribution is a H{\"o}lder function of exponent $h$ at a given 
point, the distribution is H{\"o}lder of exponent $h-n$ at the same point.

\indent 
Since the $\delta(x)$ distribution vanishes in $x\neq 0$ the
only non-trivial point is $x=0$.  Since the $\delta$ is the derivative
of $\theta(x)$ and this has H{\"o}lder exponent $0$ at $x=0$, one
concludes that the \underline{$\delta(x)$ has H{\"o}lder exponent $-1$}
\noindent
\underline{at $x=0$}

\end{enumerate}

\newpage

\clearpage

\clearpage

\begin{center}
\vspace*{-2cm}
\hbox{
	\makebox[0.5cm]{{\Large $\ln\langle \epsilon_r^3 \rangle$}}
	\makebox[8.5cm]{}
}
\vspace*{0.2cm}
\hbox{
	\makebox[0.5cm]{{\Large $\ln\langle \epsilon_r^5 \rangle$}}
	\makebox[8.5cm]{
		\leavevmode
		\epsfxsize=8.5cm
		\epsfysize=5cm
		\epsfbox[70 72 410 280]{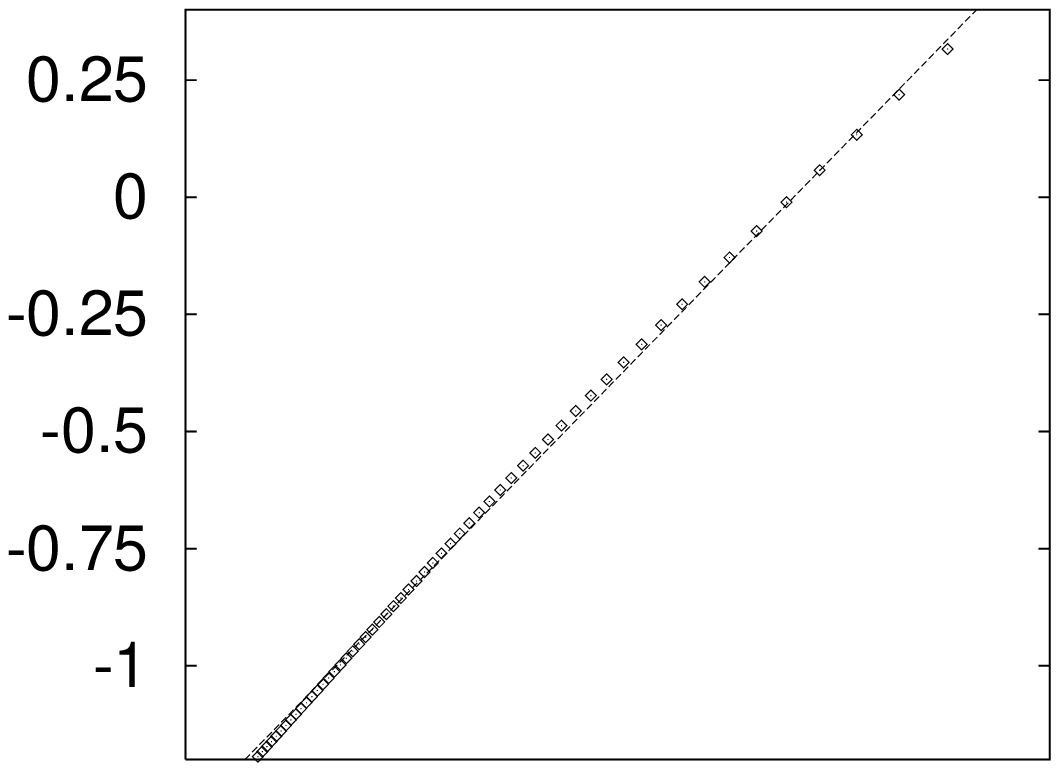}
	}
}
\hbox{
	\makebox[0.5cm]{{\Large $\ln\langle \epsilon_r^7 \rangle$}}
	\makebox[8.5cm]{
		\leavevmode
		\epsfxsize=8.5cm
		\epsfysize=5cm
		\epsfbox[70 72 410 280]{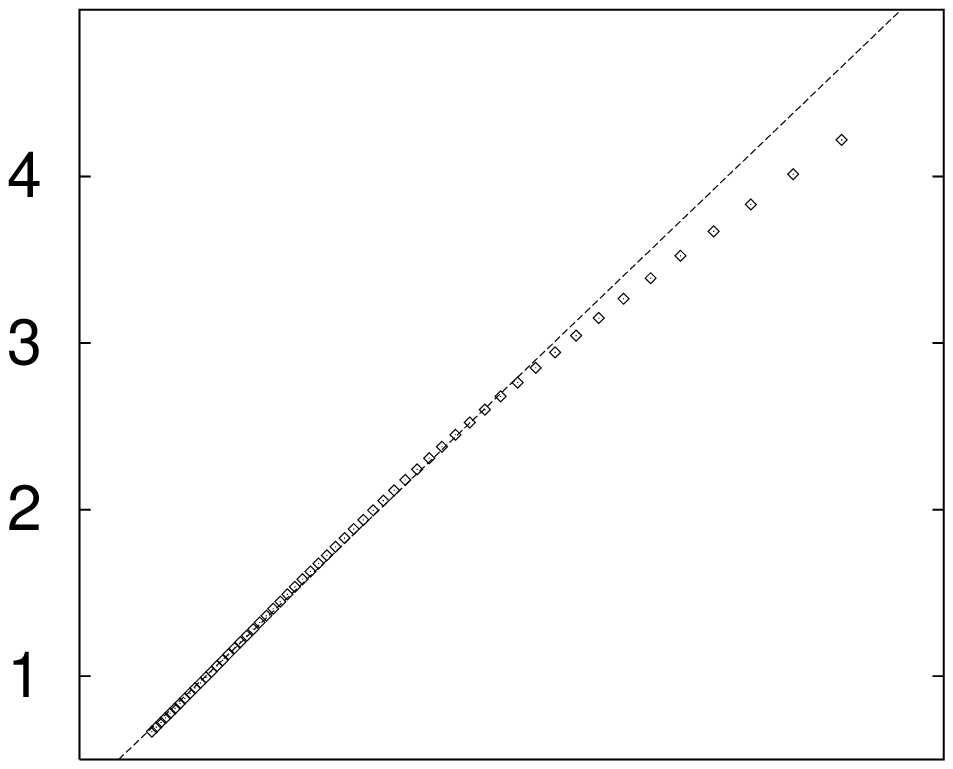}
	}
}
\hbox{
	\makebox[0.5cm]{}
	\makebox[8.5cm]{
		\leavevmode
		\epsfxsize=8.5cm
		\epsfysize=5cm
		\epsfbox[70 72 410 280]{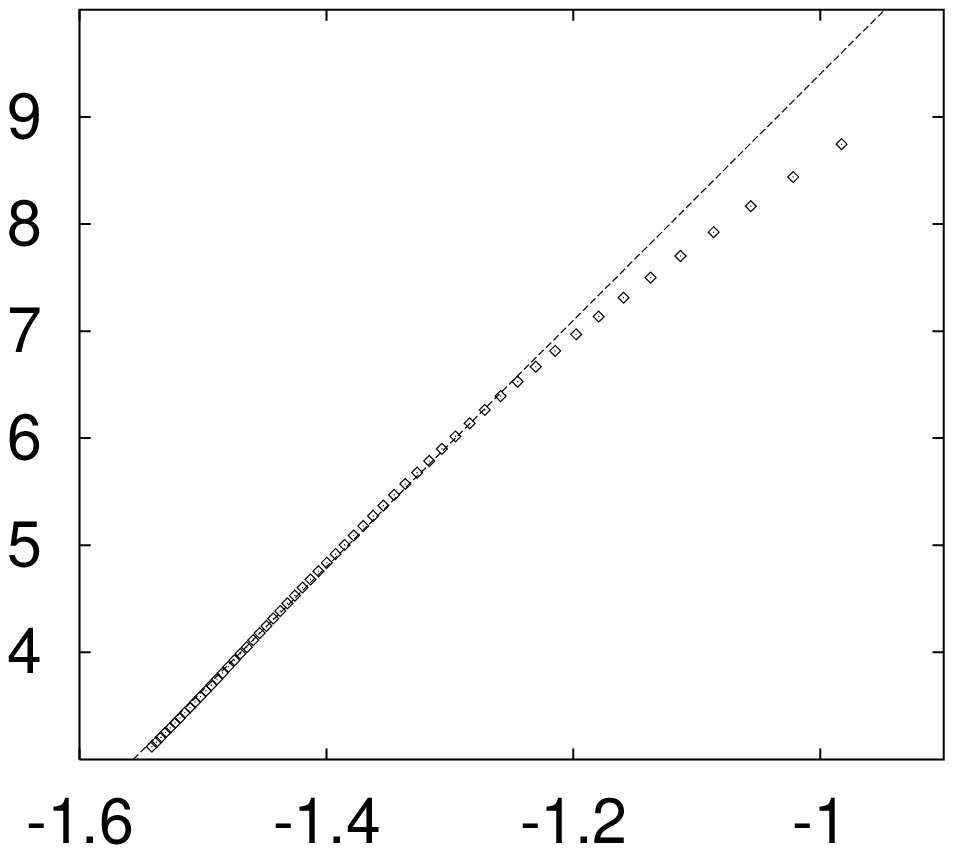}
	}
}
\vspace*{0.3cm}
\hbox{
	\makebox[0.5cm]{}
	\makebox[8.5cm]{{\Large $\ln\langle \epsilon_r^2\rangle$}}
}
\end{center}
\begin{figure}[htb]
\caption{
Test of ESS for the bidimensional edge content $\epsilon_r$.
The graphs represent the logarithm of its moments of order 3, 5 and 7
versus the logarithm of the second order moment, for distances $r=4$
to $r=64$ pixels, that is, about an order of magnitude smaller than 
the image size. This upper bound is needed both to consider values of $r$
small enough compared to the size of the images and
to be able to use moments of high order as $p=7$. The data corresponds to
the second image ensemble. According to eq.~(\ref{eq:ESS}) these graphs
should be straight lines; the best linear fits are also
represented. Except for lower cut-off effects at very small distances,
the scaling law is well verified.  
}
\label{fig:ESS-epsilon}
\end{figure}

\clearpage

\begin{center}
\vspace*{-2cm}
\hbox{
	\hspace*{-1cm}
	\makebox[1cm]{{\Large $\ln\langle \left[ T^r_{\Psi} d\mu \right]^3\rangle$}}
	\makebox[9.5cm]{{\Large \hspace*{2cm}$a$\hspace*{3.5cm}$b$}}
}
\vspace*{0.5cm}
\hbox{
	\hspace*{-1cm}
	\makebox[1cm]{{\Large $\ln\langle \left[ T^r_{\Psi} d\mu \right]^5\rangle$}}
	\makebox[9.5cm]{
		\leavevmode
		\epsfxsize=4.75cm
		\epsfysize=4.5cm
		\epsfbox[65 78 312 274]{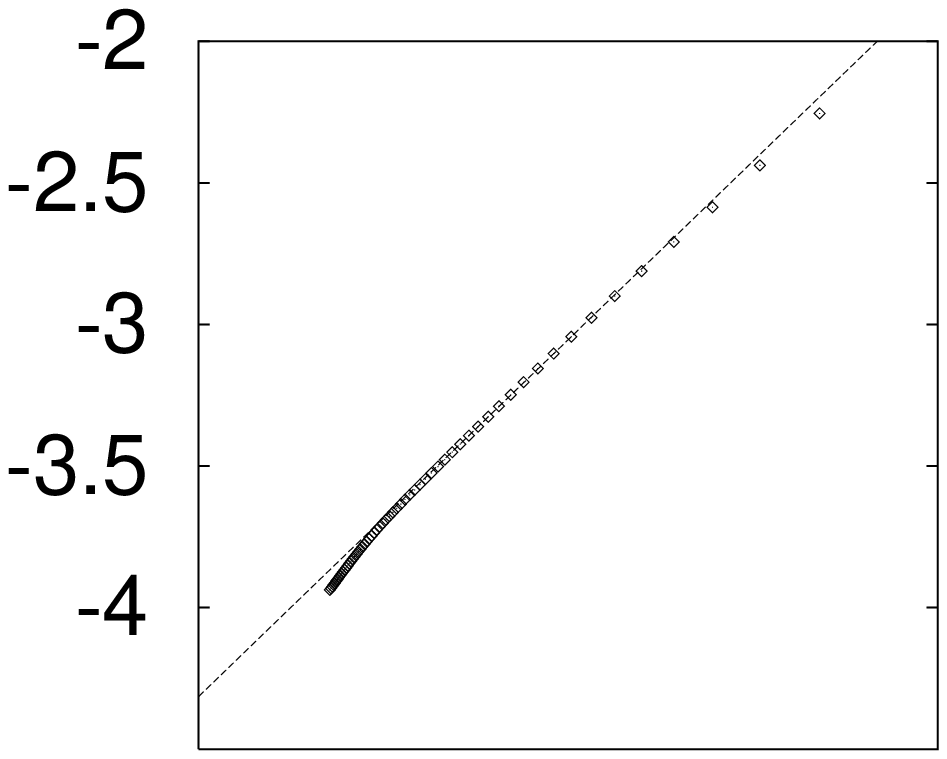}
		\epsfxsize=4.75cm
		\epsfysize=4.5cm
		\epsfbox[93 78 360 274]{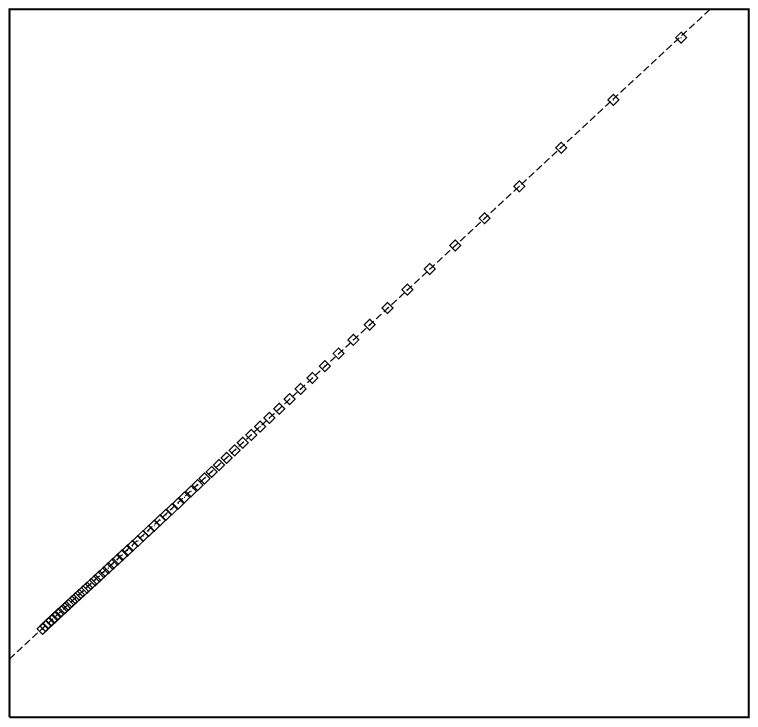}
	}
}
\hbox{
	\hspace*{-1cm}
	\makebox[1cm]{{\Large $\ln\langle \left[ T^r_{\Psi} d\mu \right]^7\rangle$}}
	\makebox[9.5cm]{
		\leavevmode
		\epsfxsize=4.75cm
		\epsfysize=4.5cm
		\epsfbox[65 78 312 274]{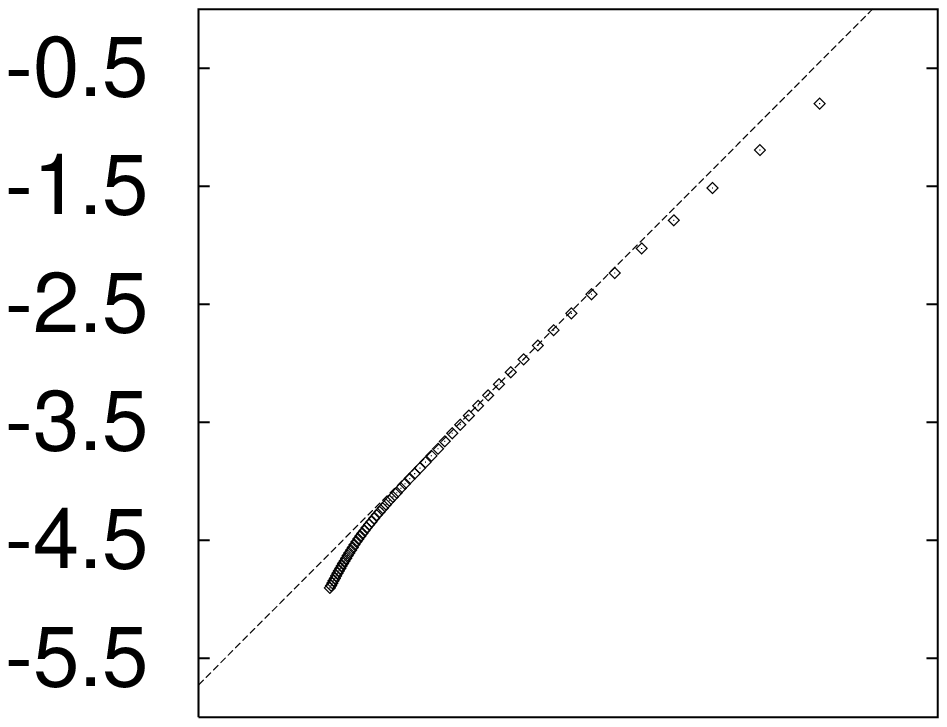}
		\epsfxsize=4.75cm
		\epsfysize=4.5cm
		\epsfbox[93 78 360 274]{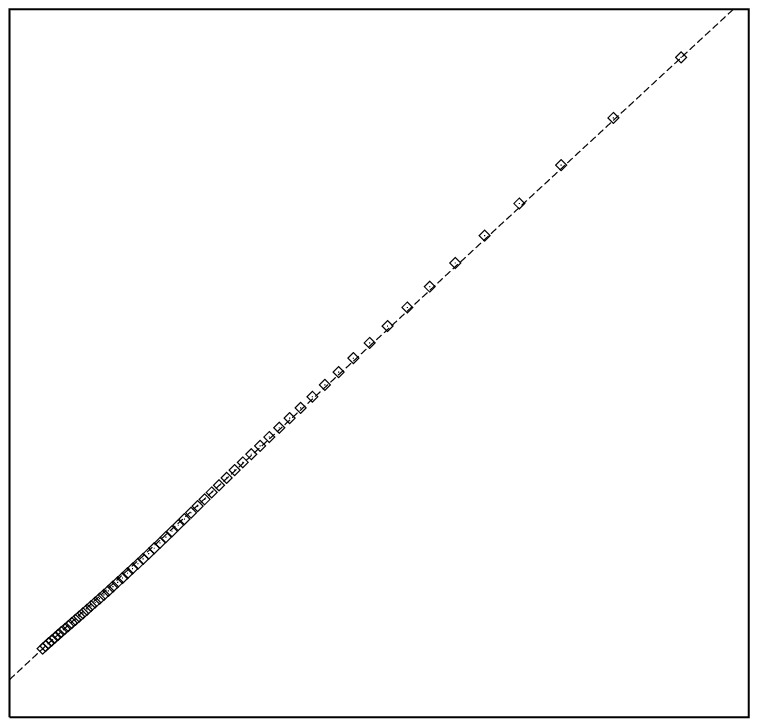}
	}
}
\hbox{
	\hspace*{-1cm}
	\makebox[1cm]{}
	\makebox[9.5cm]{
		\leavevmode
		\epsfxsize=4.75cm
		\epsfysize=4.5cm
		\epsfbox[65 78 312 274]{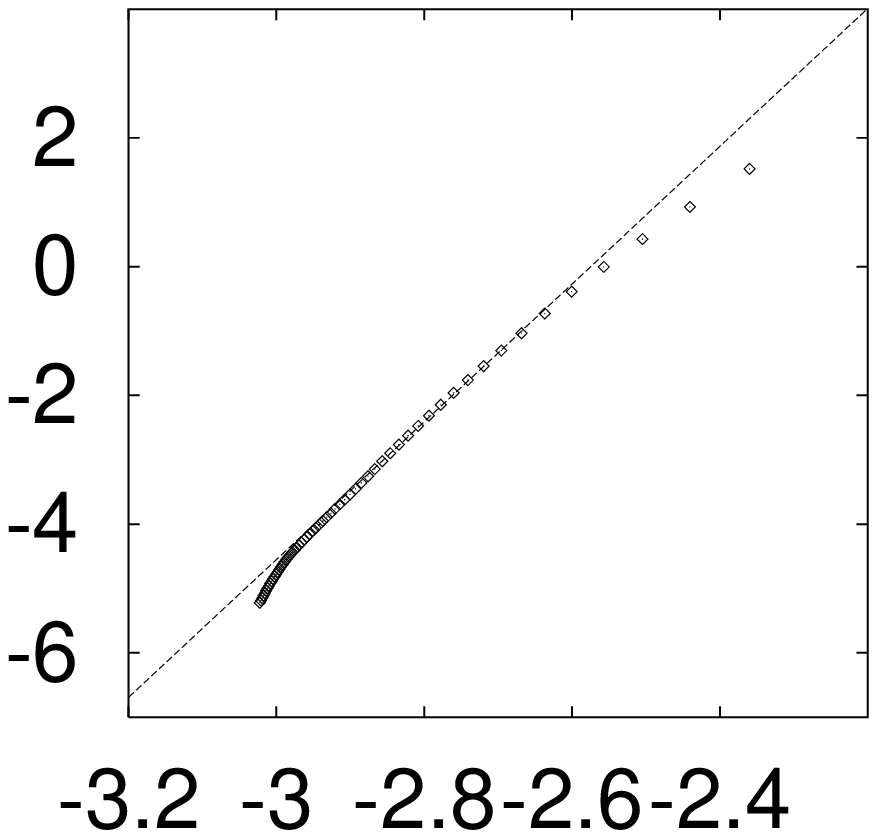}
		\epsfxsize=4.75cm
		\epsfysize=4.5cm
		\epsfbox[93 78 360 274]{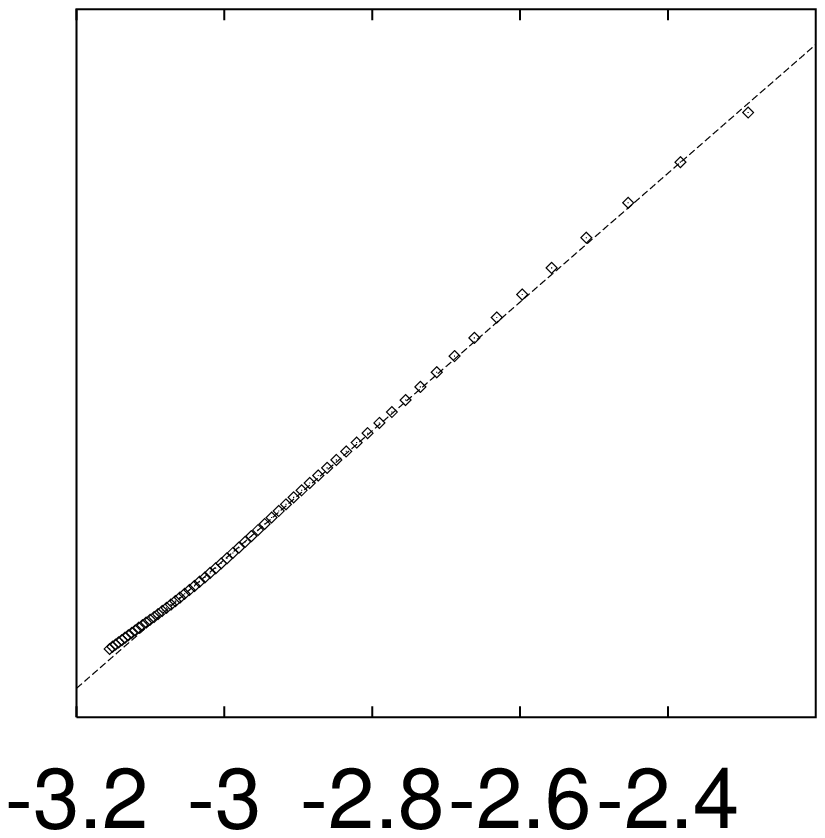}
	}
}
\vspace*{0.5cm}
\hbox{
	\hspace*{-1cm}
	\makebox[1cm]{}
	\makebox[9.5cm]{\hspace*{1.75cm}{\Large $\ln\langle \left[
T^r_{\Psi} d\mu\right]^2\rangle$\hspace*{1.5cm} $\ln\langle \left[
T^r_{\Psi} d\mu \right]^2\rangle$}}
}
\end{center}
\begin{figure}[htb]
\caption{
Test of ESS for {\bf a)} $T^r_{\Psi} d\mu ^h$ and {\bf b)} $T^r_{\Psi} d\mu ^v$. The 
graphs represent the logarithm of the moments of order 
3, 5 and 7 for both variables versus the logarithm of the corresponding 
second order moment. The scales range from $r=4$ to $r=64$ pixels. The 
first dataset was used in this computation. The corresponding best linear 
fits are also represented. Except for lower and upper cut-off effects at 
small and large distances, the scaling law verifies well. Although the
test is not shown here, SS also holds for these variables.
}
\label{fig:ESS-grad_wv}
\end{figure}

\clearpage

\clearpage

\begin{center}
\hbox{
	\hspace*{-1.5cm}
	\makebox[1.7cm]{{\large $\ln\langle \left[\delta_{\vec{r}}C\right]^3\rangle$}}
	\makebox[10.5cm]{{\large $a$\hspace*{5cm}$b$}}
	\makebox[1.5cm]{{\large $\ln\langle\left[T^{\vec{r}}_{\Psi}C\right]^3\rangle$}}
}
\vspace*{0.2cm}
\hbox{
	\hspace*{-1.5cm}
	\makebox[1.7cm]{{\large $\ln\langle \left[\delta_{\vec{r}}C\right]^5\rangle$}}
	\makebox[10.5cm]{
		\leavevmode
		\epsfxsize=5cm
		\epsfbox[100 80 410 284]{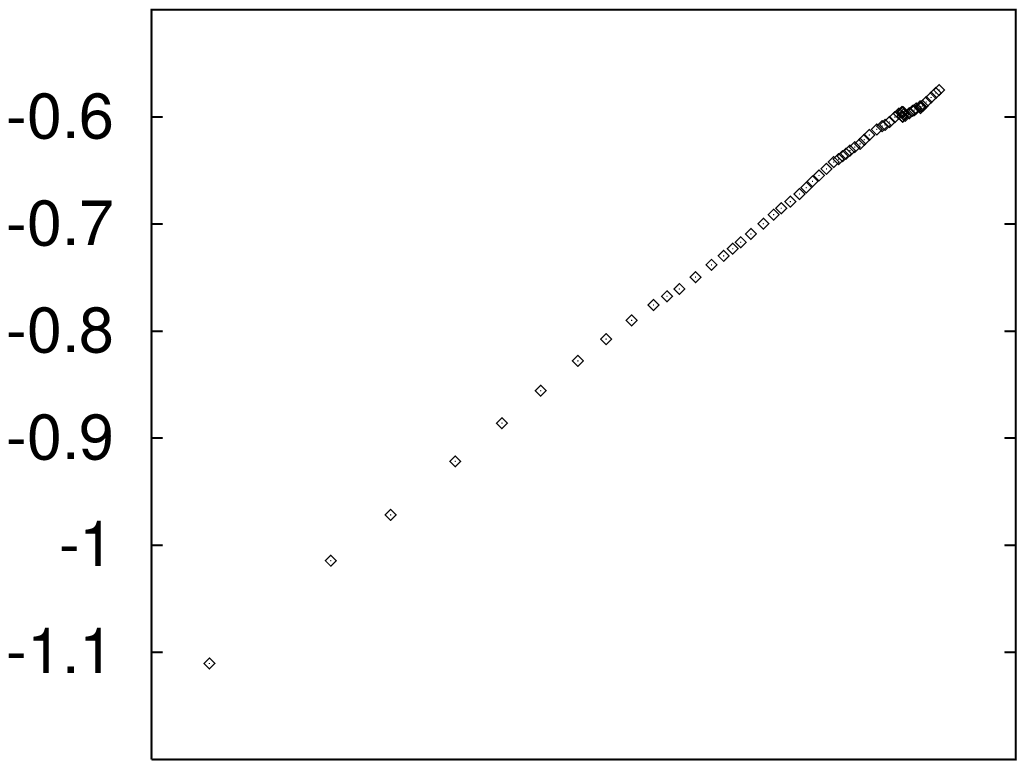}
		\hspace*{0.1cm}
		\epsfxsize=5cm
		\epsfbox[100 80 410 284]{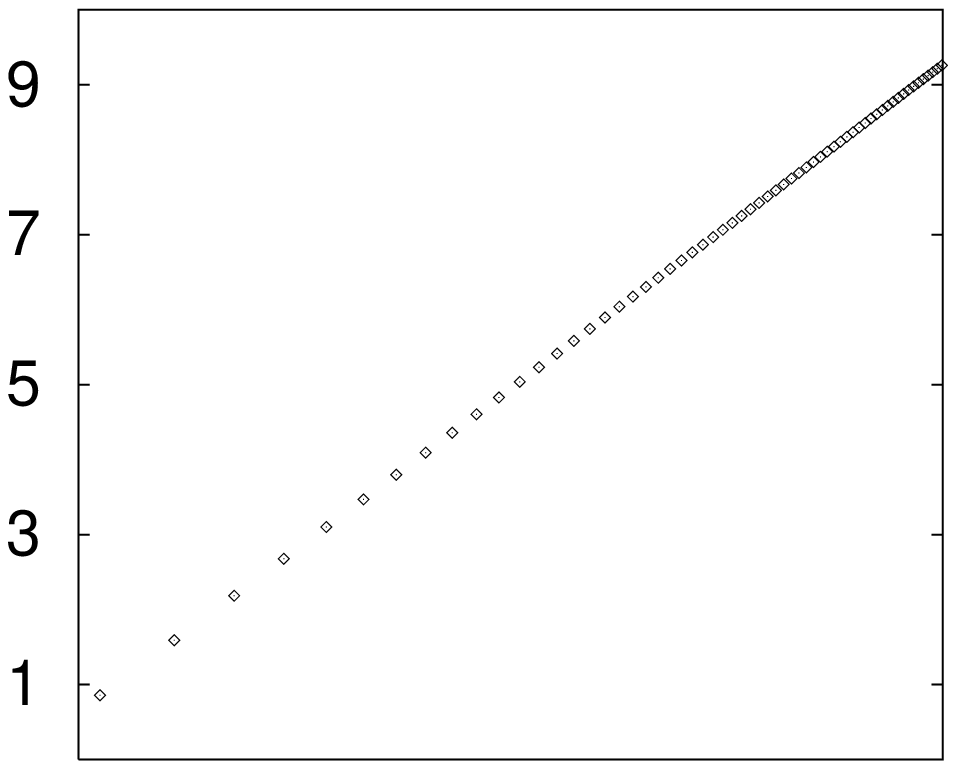}
	}
	\makebox[1.5cm]{{\large $\ln\langle\left[T^{\vec{r}}_{\Psi}C\right]^5\rangle$}}
}
\hbox{
	\hspace*{-1.5cm}
	\makebox[1.7cm]{{\large $\ln\langle \left[\delta_{\vec{r}}C\right]^7\rangle$}}
	\makebox[10.5cm]{
		\leavevmode
		\epsfxsize=5cm
		\epsfbox[100 80 410 284]{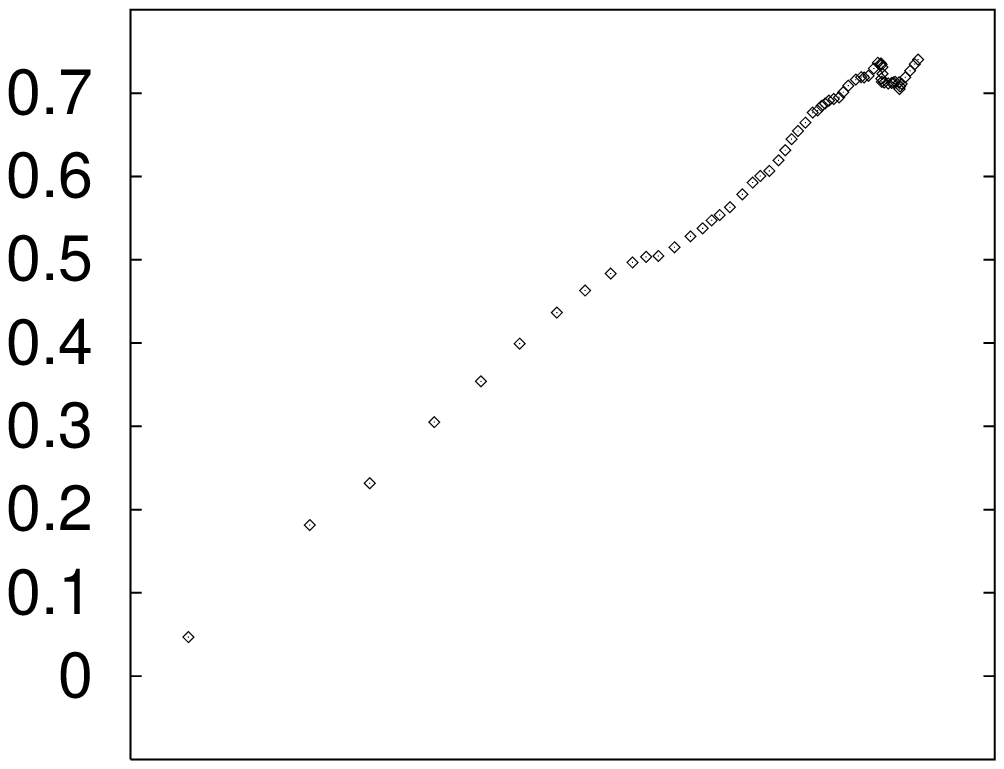}
		\hspace*{0.1cm}
		\epsfxsize=5cm
		\epsfbox[100 80 410 284]{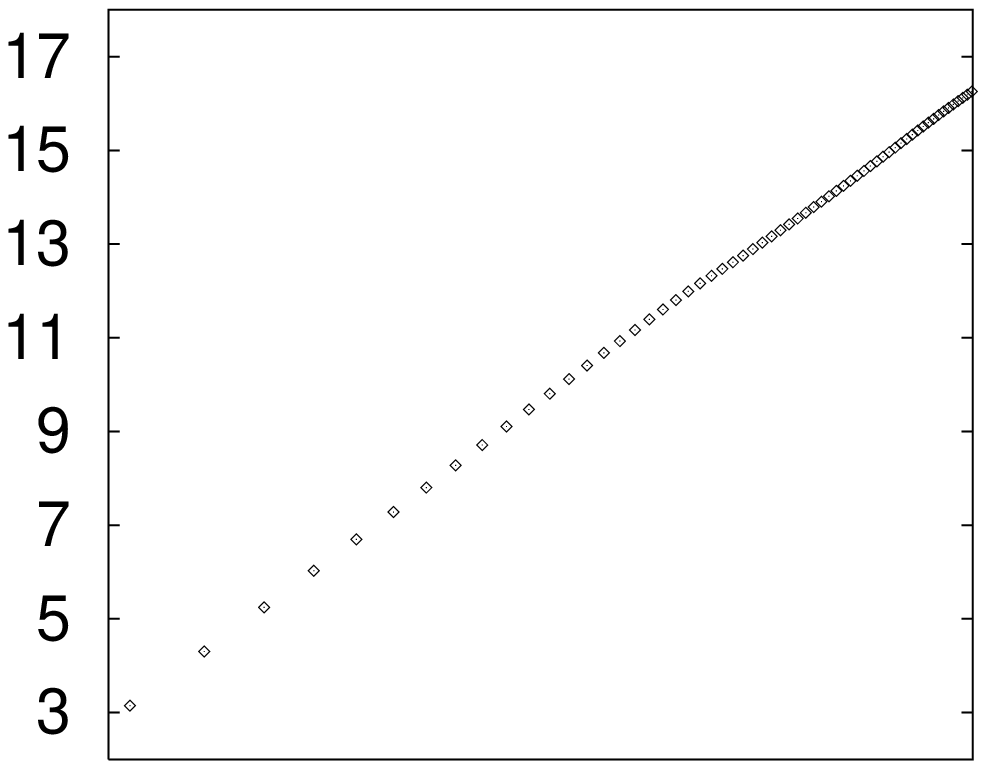}
	}
	\makebox[1.5cm]{{\large $\ln\langle\left[T^{\vec{r}}_{\Psi}C\right]^7\rangle$}}
}
\hbox{
	\hspace*{-1.5cm}
	\makebox[1.7cm]{}
	\makebox[10.5cm]{
		\leavevmode
		\epsfxsize=5cm
		\epsfbox[100 80 410 284]{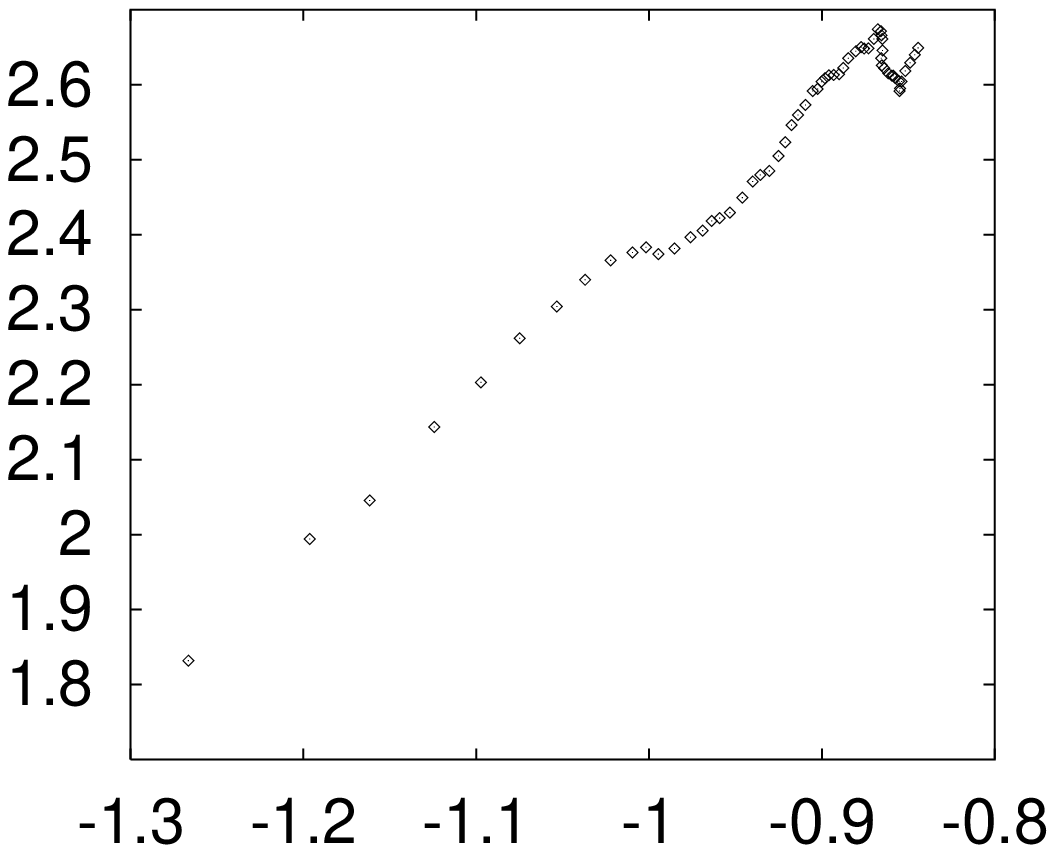}
		\hspace*{0.1cm}
		\epsfxsize=5cm
		\epsfbox[100 80 410 284]{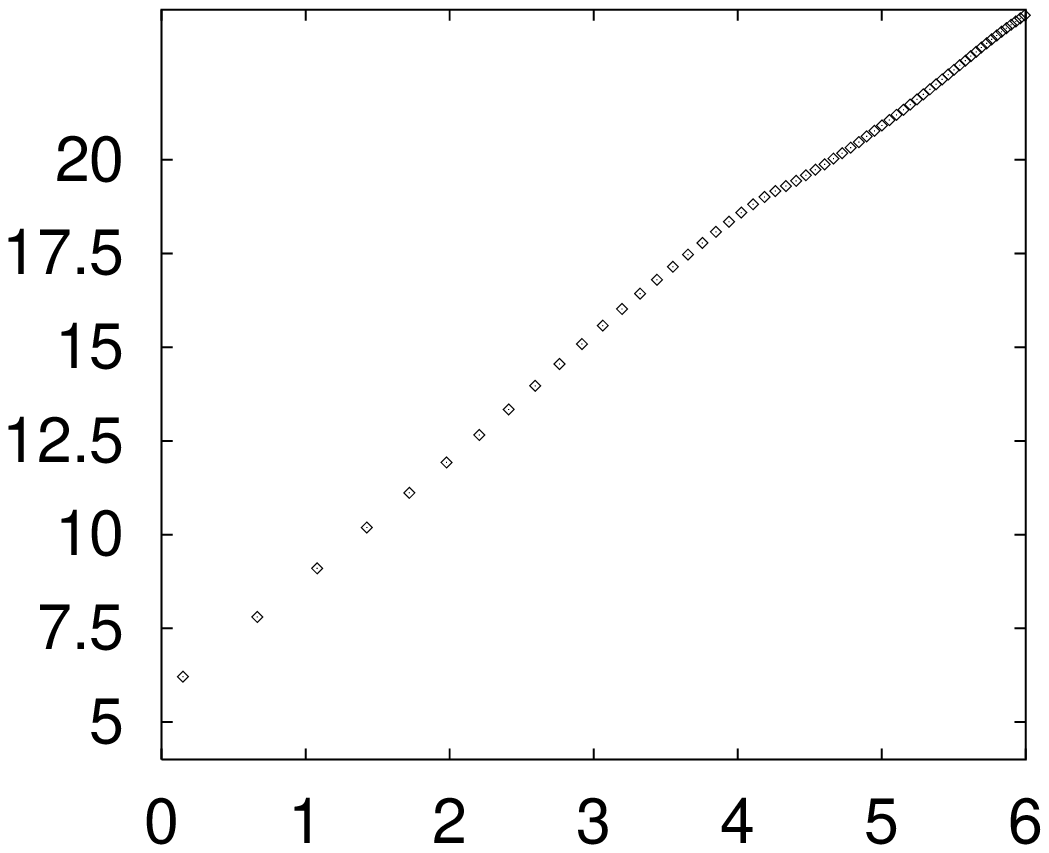}
	}
	\makebox[1.5cm]{}
}
\vspace*{0.3cm}
\hbox{
	\hspace*{-1.5cm}
	\makebox[1.7cm]{}
	\makebox[10.5cm]{{\large $\ln\langle \left[\delta_{\vec{r}}C\right]^2\rangle$
	\hspace*{3cm}$\ln\langle\left[T^{\vec{r}}_{\Psi}C\right]^2\rangle$}}
	\makebox[1.5cm]{}
}
\end{center}

\clearpage

\begin{center}
\hbox{
	\hspace*{-1.5cm}
	\makebox[1.5cm]{{\large $\ln\langle \left[\delta_{\vec{r}} 
|\nabla C|\right]^3\rangle$}}
	\makebox[9.5cm]{{\large \hspace*{2cm}$a$\hspace*{4.5cm}$b$}}
}
\vspace*{0.5cm}
\hbox{
	\hspace*{-1.5cm}
	\makebox[1.5cm]{{\large $\ln\langle \left[\delta_{\vec{r}} 
|\nabla C|\right]^5\rangle$}}
	\makebox[9.5cm]{
		\leavevmode
		\epsfxsize=4.75cm
		\epsfbox[80 72 328 280]{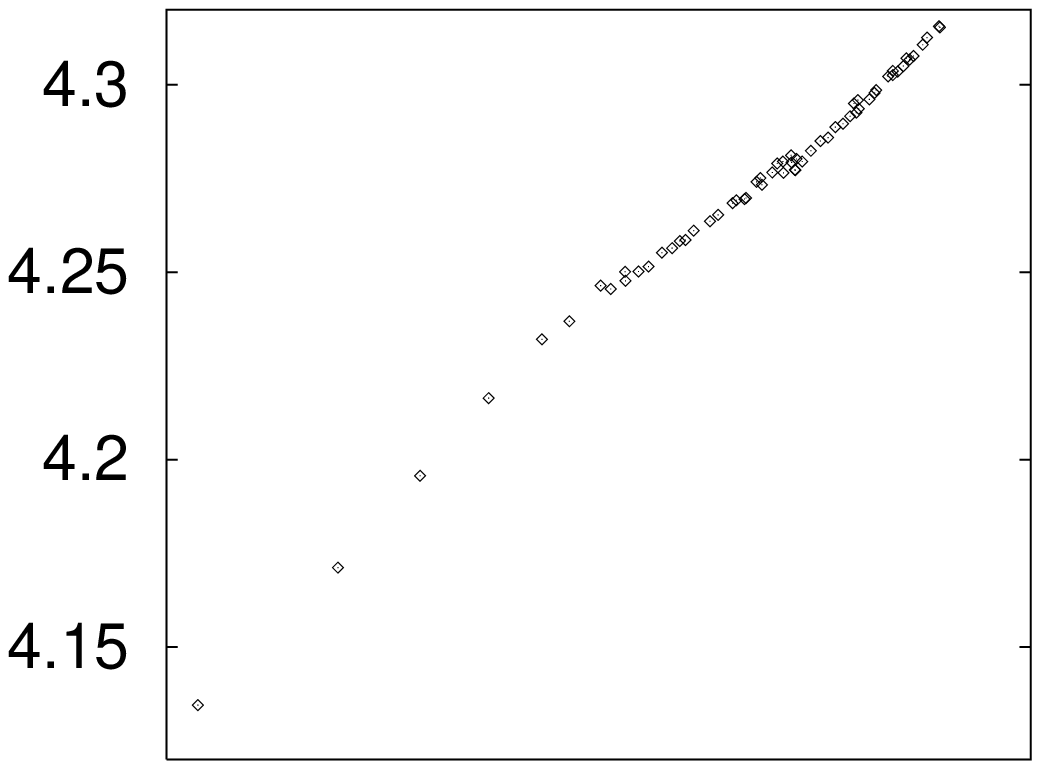}
		\epsfxsize=4.75cm
		\epsfbox[80 72 328 280]{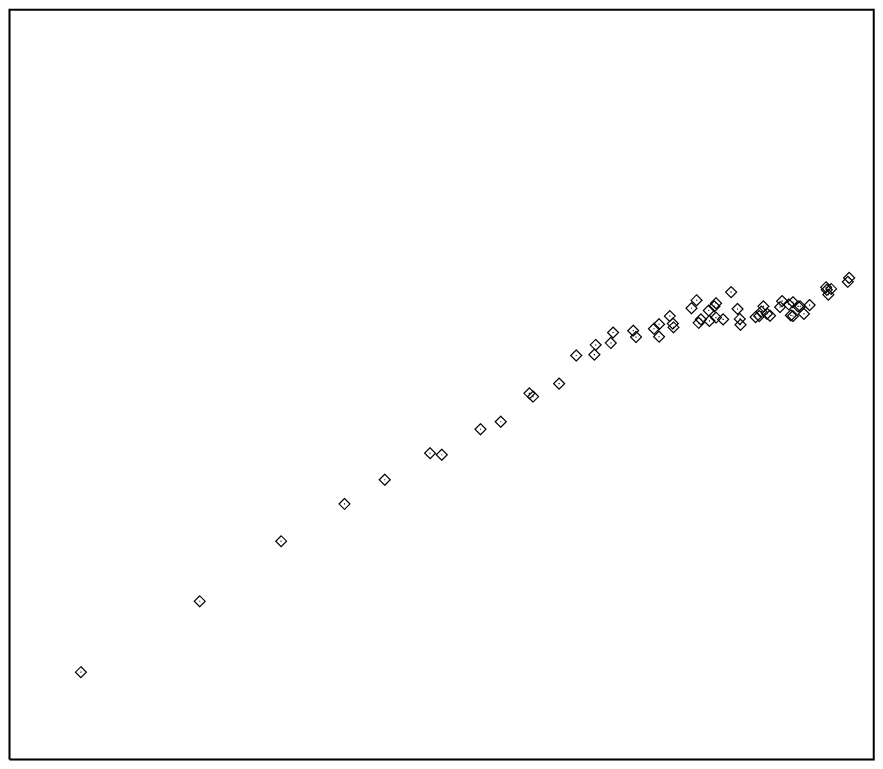}
	}
}
\hbox{
	\hspace*{-1.5cm}
	\makebox[1.5cm]{{\large $\ln\langle \left[\delta_{\vec{r}} 
|\nabla C|\right]^7\rangle$}}
	\makebox[9.5cm]{
		\leavevmode
		\epsfxsize=4.75cm
		\epsfbox[80 72 328 280]{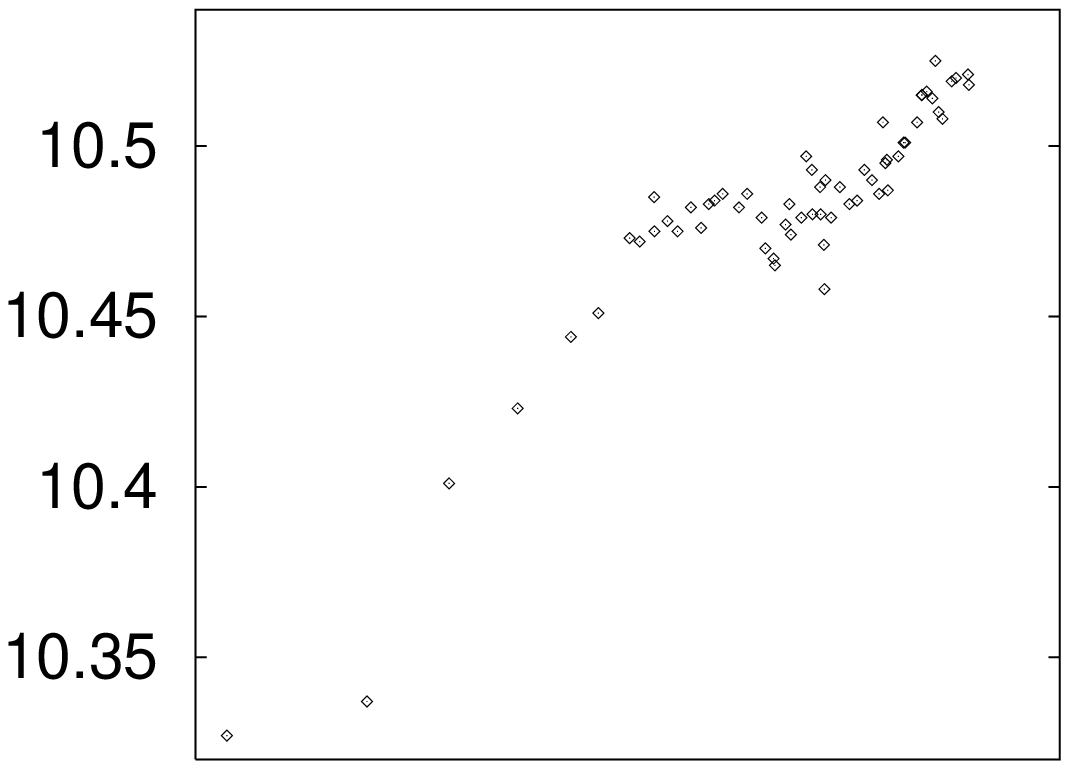}
		\epsfxsize=4.75cm
		\epsfbox[80 72 328 280]{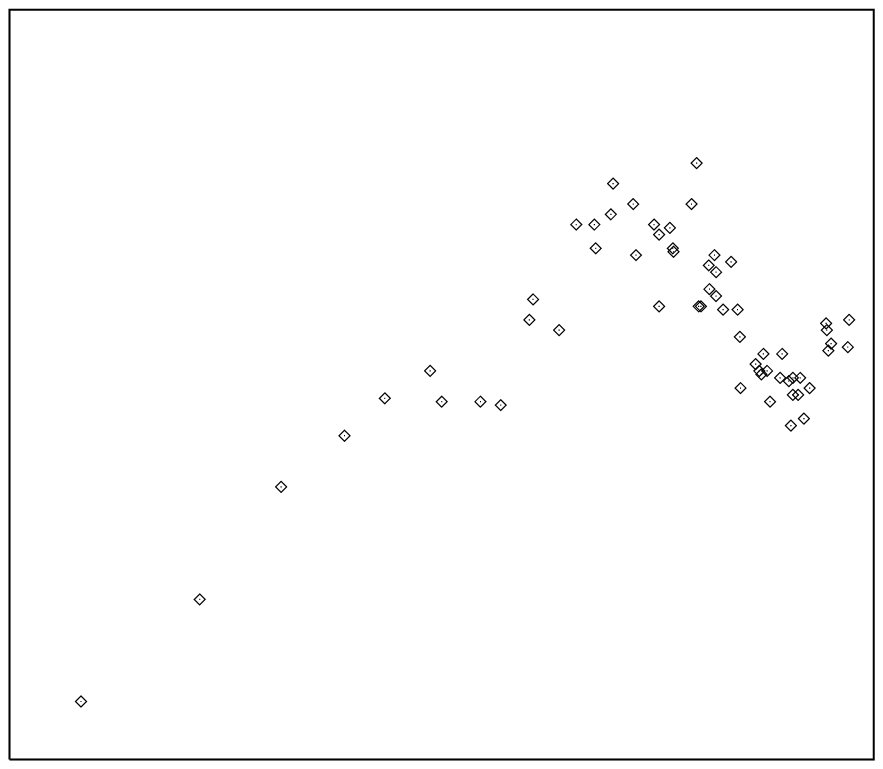}
	}
}
\hbox{
	\hspace*{-1.5cm}
	\makebox[1.5cm]{}
	\makebox[9.5cm]{
		\leavevmode
		\epsfxsize=4.75cm
		\epsfbox[80 72 328 280]{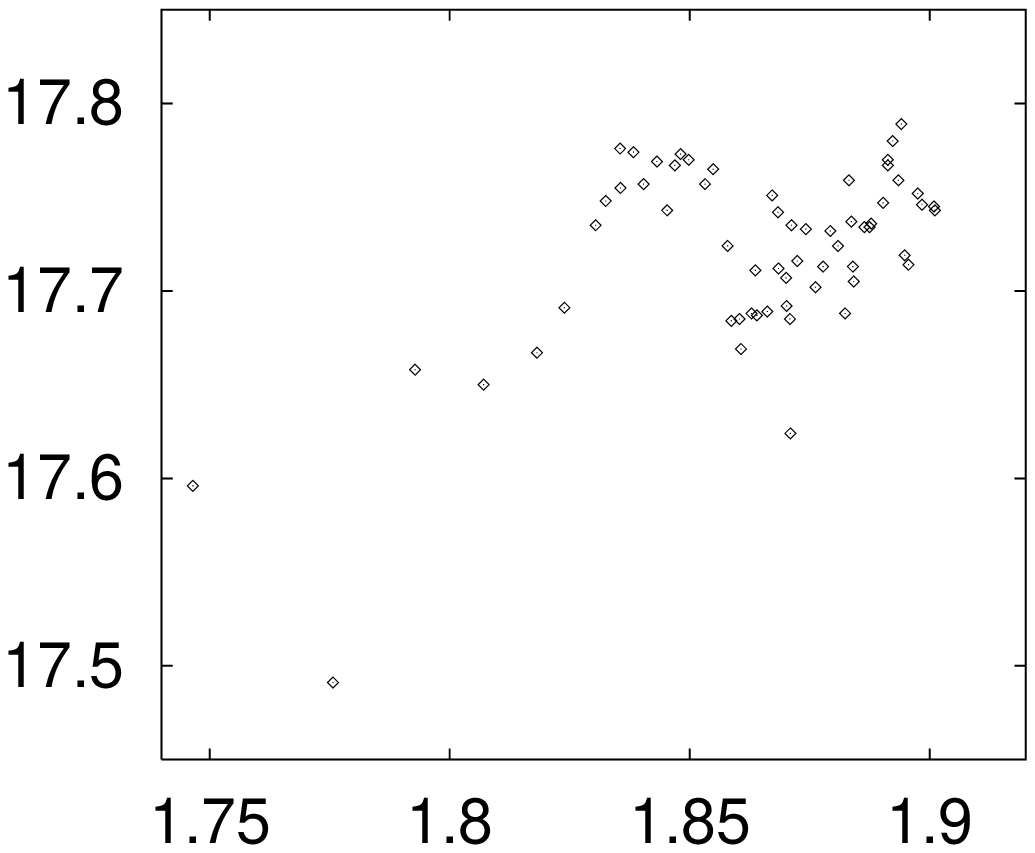}
		\epsfxsize=4.75cm
		\epsfbox[80 72 328 280]{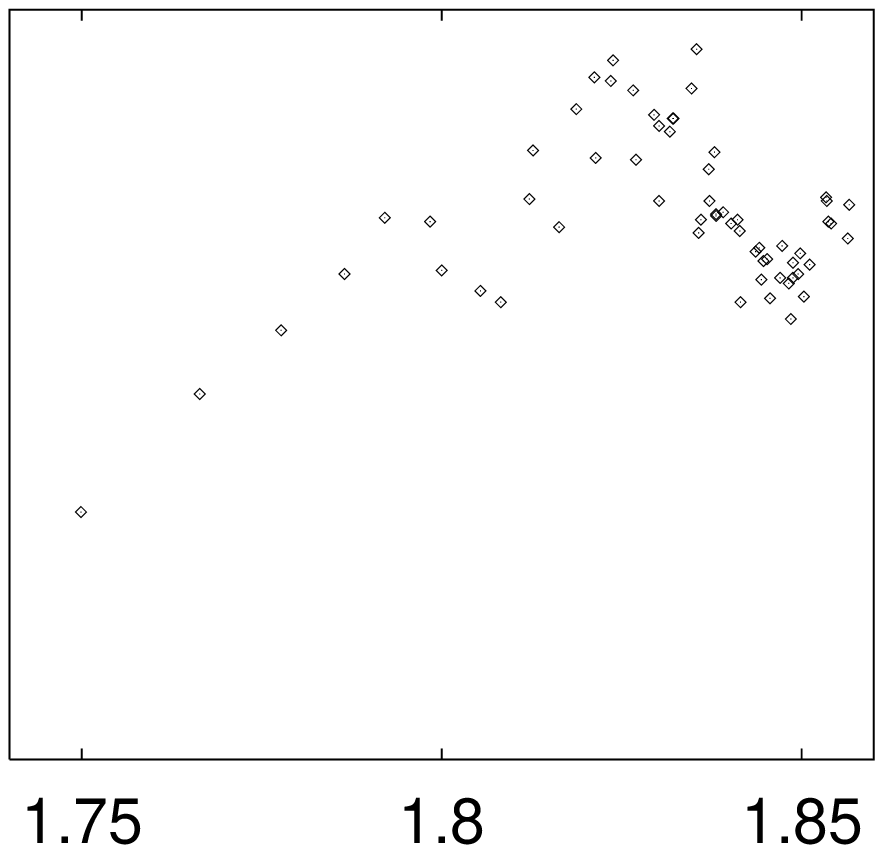}
	}
}
\vspace*{0.5cm}
\hbox{
	\hspace*{-1.5cm}
	\makebox[1.5cm]{}
	\makebox[9.5cm]{\hspace*{2cm}{\large $\ln\langle
\left[\delta_{\vec{r}}  |\nabla
C|\right]^2\rangle$\hspace*{2cm}$\ln\langle \left[\delta_{\vec{r}} 
|\nabla C|\right]^2\rangle$}}
}
\end{center}
\begin{figure}[htb]
\caption{
Test of ESS for {\bf a)} horizontal linear increment of $|\nabla C|$,
$\delta_{r\vec{i}}|\nabla C|$ and {\bf b)} vertical linear increment of
the same function, $\delta_{r\vec{j}}|\nabla C|$. The first dataset
was used and the scales were taken ranging from $r=4$ to $r=64$
pixels. Again, these variables do not have ESS.  The corresponding
wavelet projections are shown in Fig. \ref{fig:ESS-grad_wv} where one
can see that ESS holds.
} 
\label{fig:ESS-gradLI}
\end{figure}

\clearpage

\end{document}